\documentclass[10pt, conference, compsocconf]{IEEEtran}

\ifCLASSINFOpdf
\else
\fi

\usepackage{cite}
\usepackage{amsmath,amssymb,amsfonts}
\usepackage{booktabs}
\usepackage{multirow}
\usepackage{algorithm}
\usepackage[noend]{algpseudocode}
\usepackage[caption=false]{subfig}
\usepackage{graphicx}
\usepackage{textcomp}
\usepackage{xcolor}

\usepackage{tabularx}
\usepackage{soul}
\usepackage{lipsum}
\usepackage{diagbox}

\usepackage{pifont}
\newcommand{\cmark}{\ding{51}}
\newcommand{\xmark}{\ding{55}}

\usepackage[normalem]{ulem}

\algnewcommand\algorithmicforeach{\textbf{for each}}
\algdef{S}[FOR]{ForEach}[1]{\algorithmicforeach\ #1\ \algorithmicdo}

\algtext*{EndWhile}
\algtext*{EndIf}
\algtext*{EndFor}

\renewcommand{\baselinestretch}{0.992}

\begin{document}

\title{SCRAMBLE: The State, Connectivity and Routing Augmentation Model \\ for Building Logic Encryption}

\author{
    \IEEEauthorblockN{
        Hadi~Mardani~Kamali\IEEEauthorrefmark{1}, 
        Kimia~Zamiri~Azar\IEEEauthorrefmark{1},
        Houman~Homayoun\IEEEauthorrefmark{2}, 
        Avesta~Sasan\IEEEauthorrefmark{1}}
    \IEEEauthorblockA{
        \IEEEauthorrefmark{1}Department of ECE, George Mason University, {\normalfont E-mail: \{hmardani, kzamiria, asasan\}@gmu.edu} \\          
        \IEEEauthorrefmark{2}Department of ECE, University of California, Davis, {\normalfont E-mail: hhomayoun@ucdavis.edu} \\
    }
}

\maketitle

\begin{abstract}

In this paper, we introduce SCRAMBLE, as a novel logic locking solution for sequential circuits while the access to the scan chain is restricted. The SCRAMBLE could be used to lock an FSM by hiding its state transition graph (STG) among a large number of key-controlled false transitions. Also, it could be used to lock sequential circuits (sequential datapath) by hiding the timing paths' connectivity among a large number of key-controlled false connections. Besides, the structure of SCRAMBLE allows us to engage this scheme as a new scan chain locking solution by hiding the correct scan chain sequence among a large number of the key-controlled false sequences. We demonstrate that the proposed scheme resists against both (1) the 2-stage attacks on FSM, and (2) SAT attacks integrated with unrolling as well as bounded-model-checking. We have discussed two variants of SCRAMBLE: (I) Connectivity SCRAMBLE (SCRAMBLE-C), and (b) Logic SCRAMBLE (SCRAMBLE-L). The SCRAMBLE-C relies on the SAT-hard and key-controlled modules that are built using near non-blocking logarithmic switching networks. The SCRAMBLE-L uses input multiplexing techniques to hide a part of the FSM in a memory. In the result section, we describe the effectiveness of each variant against state-of-the-art attacks.

\end{abstract}


\IEEEpeerreviewmaketitle

\section{Introduction}

Due to the ever-increasing cost of IC manufacturing, many design houses have been forced to become fabless. Outsourcing the manufacturing stages, including fabrication/testing/packaging, to potentially untrusted entities raises multiple forms of security threats such as IC overproduction, Trojan insertion, reverse engineering (RE), intellectual property (IP) theft, and counterfeiting \cite{rostami2014primer, tehranipoor2017invasion}. 

Among many active and passive \emph{design-for-trust} (DfT) techniques, logic locking \cite{roy2010ending, rajendran2012security}, as a proactive solution, hides the functionality of the netlist by inserting additional programmable gates (key gates), whose programming values (key values) are stored in tamper-proof memories. However, the strength of logic locking solutions was seriously challenged in recent years by the Boolean satisfiability (SAT) attack \cite{subramanyan2015evaluating, el2015integrated}.

Although the SAT attack (and many of its derivatives) only works on combinational circuits \cite{zamiri2019threats}\nocite{azar2019smt}, having access to the scan chain allows an adversary to apply the SAT attack on each combinational logic part of sequential circuits separately. Accordingly, the adversary can target sequential circuits using the SAT attack. Hence, few recent studies investigated the possibility of restricting the scan chain using scan chain locking/blocking \cite{karmakar2018encrypt, guin2018robust, wang2018secure}\nocite{roshanisefat2020dfssd,kamali2020designing}. Also, considering that the access to the scan chain is restricted/locked, several studies investigated the possibility of applying the logic locking to the whole sequential circuits \cite{meade2017revisit, li2013structural},  particularly FSMs  \cite{chakraborty2009harpoon, li2013structural, meade2017revisit, koushanfar2017active, dofe2018novel, fyrbiak2018difficulty}\nocite{azar2019coma}. However, further research revealed that new attacks could be formulated for these locking solutions even while access to the scan chain is restricted/locked.  

In case of FSM locking \cite{chakraborty2009harpoon, dofe2018novel, koushanfar2017active, desai2013interlocking}, a new attack, without oracle access, denoted as \emph{\underline{\textbf{2-stage attacks on FSM}}} (2-stage) was formulated \cite{meade2017revisit, fyrbiak2018difficulty}. Also, in case of sequential (datapath) or scan chain locking \cite{meade2017revisit, li2013structural, karmakar2018encrypt, wang2018secure, guin2018robust}, a new breed of SAT-based attacks, referred as \emph{\underline{\textbf{unrolling-based SAT attack}}} (UB-SAT) as well as SAT attacks integrated with bounded-model-checking (BMC) was formulated \cite{shamsi2019kc2, el2017reverse, alrahis2019scansat}, challenging the validity of these solutions. 

To defend against UB-SAT or BMC, and to break 2-stage attacks on FSMs, we introduce a new \underline{s}tate, \underline{c}onnectivity and \underline{r}outing \underline{a}ugmentation \underline{m}odel for \underline{b}uilding \underline{l}ogic \underline{e}ncryption (SCRAMBLE). SCRAMBLE is designed to add and maximize (a) the \emph{false transitions} within STG when FSM is targeted for locking, (b) the \emph{false connections} between datapath flip flops (FFs) when sequential datapath locking is targeted, (c) the \emph{false sequences'} possibilities in scan FFs (SFFs) when the scan chain is targeted. SCRAMBLE, with 2 variants, can resist both 2-stage attacks on FSM as well as UB-SAT or BMC attacks on sequential datapath or scan chain locking.

\section{Background and Related Work} \label{background}

\subsection{FSM, Sequential Datapath, and Scan Chain Locking}

In FSM locking \cite{chakraborty2009harpoon, desai2013interlocking, koushanfar2017active, dofe2018novel}, few extra sets (modes) of states are added to the original state transition graph (STG), such as locking/authentication mode states or black holes. The traversal sequence of locking/authentication modes is the locking/authentication key, and a correct traversal that reaches the initial state of the original STG unlocks the FSM. Also, the output generated by the correct traversal of authentication states serves as a watermark. In addition to these groups, a set of studies focuses on locking the STG without adding any extra state. However, the complexity and overhead (area) of this approach is higher compared to other schemes \cite{meade2017revisit}. 

In sequential datapath locking or scan chain locking, XOR- or MUX-based locking has been added into timing paths or the scan chain. For instance, in scan chain locking solutions,  the scan chain has been blocked or locked to prevent the scan chain loading and readout (shift/load) for protecting the logic against the SAT attacks. For example, the \emph{Encrypt Flip-Flop} solution \cite{karmakar2018encrypt} adds key-programmable MUXes on the outputs of SFFs enabling the negation of the SFFs when the scan chain locking key is incorrect.

\subsection{Attacks on FSM, Sequential, and Scan Chain Locking}  \label{attacks_described}

\begin{figure}[t]
\centering
\subfloat[HARPOON \cite{chakraborty2009harpoon}]{{\includegraphics[width=0.48\columnwidth]{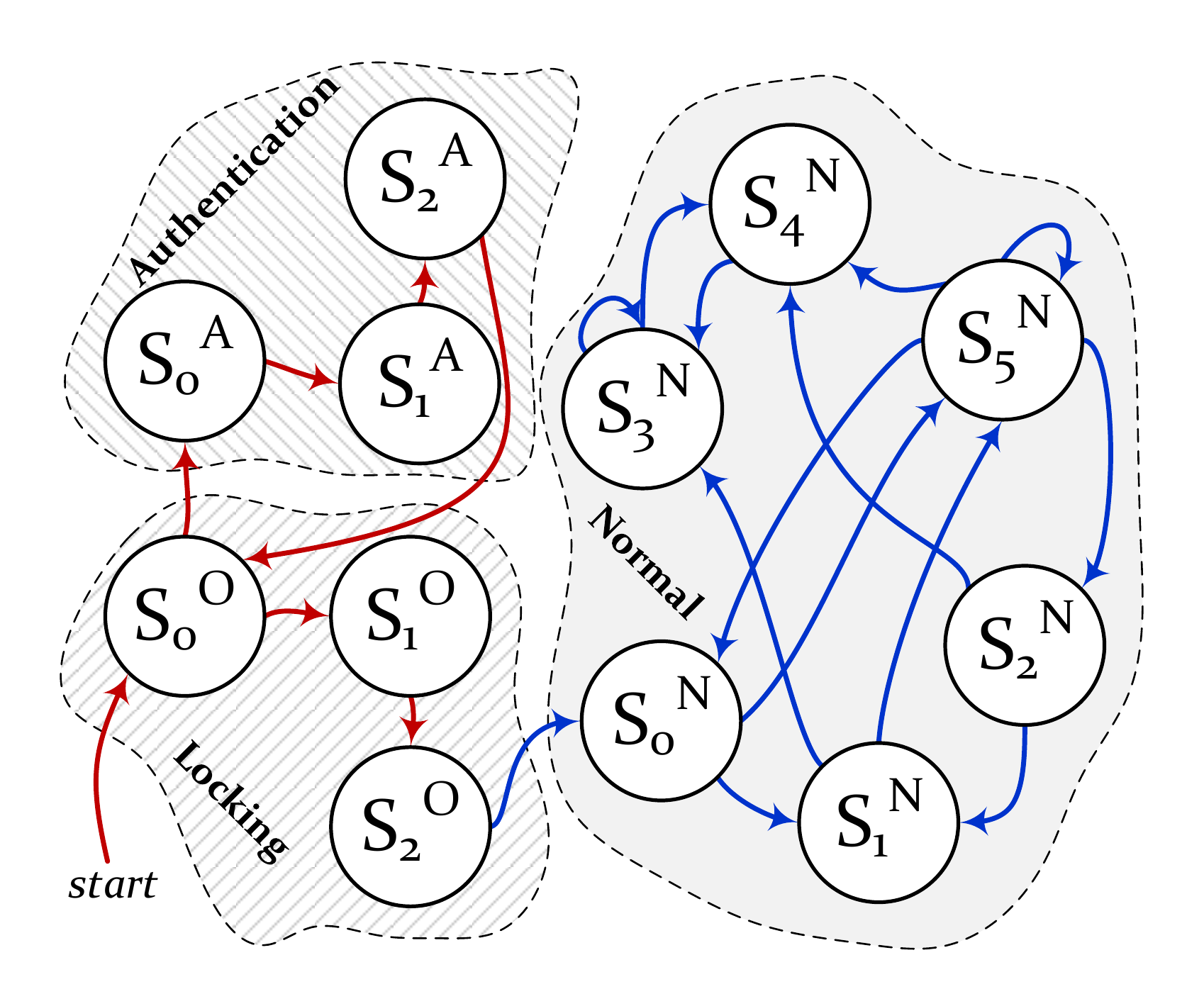} }}%
\subfloat[Dynamic State Deflection \cite{dofe2018novel}]{{\includegraphics[width=0.48\columnwidth]{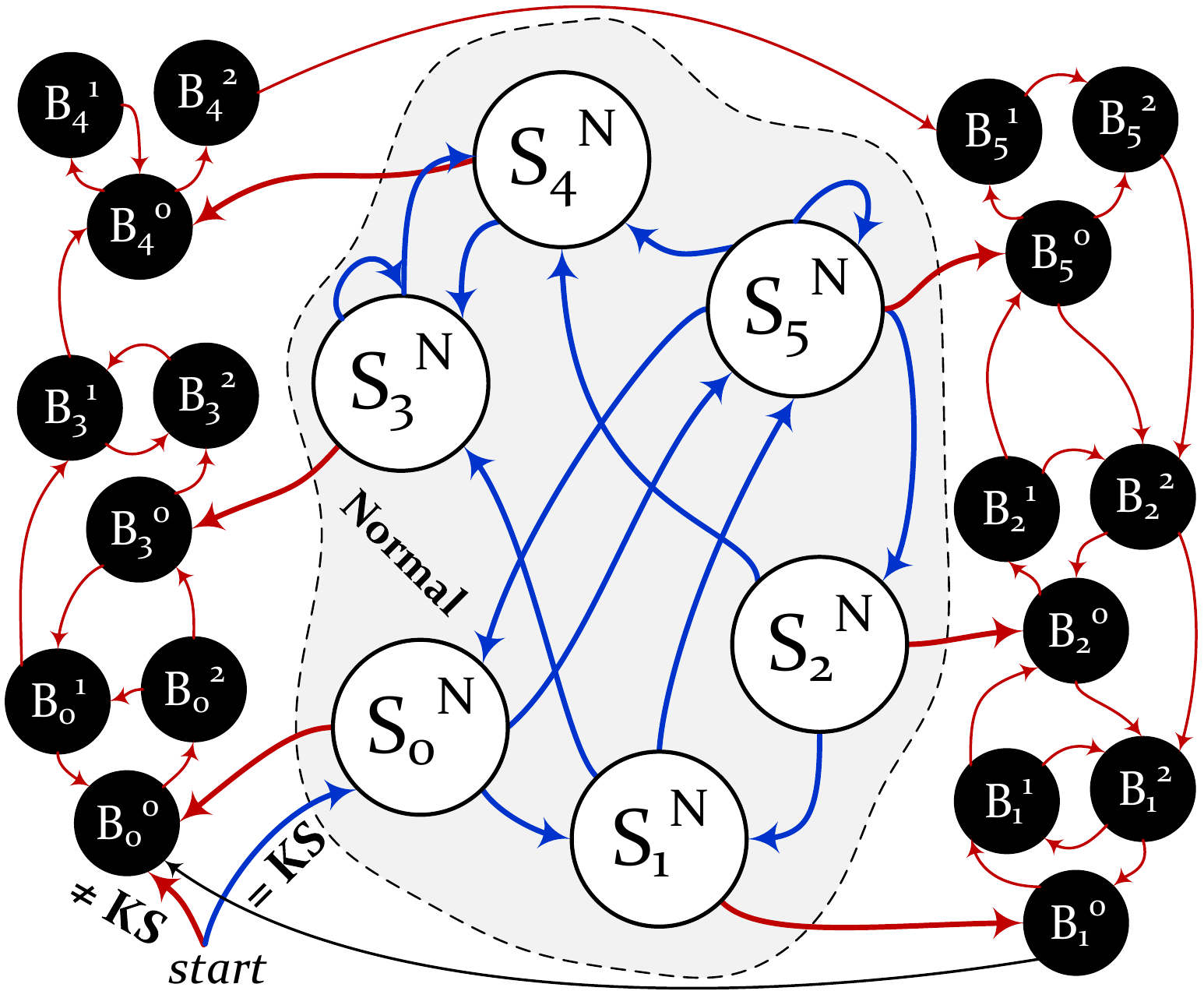} }} \\%
\subfloat[Hardware Active Metering \cite{koushanfar2017active}]{{\includegraphics[width=0.48\columnwidth]{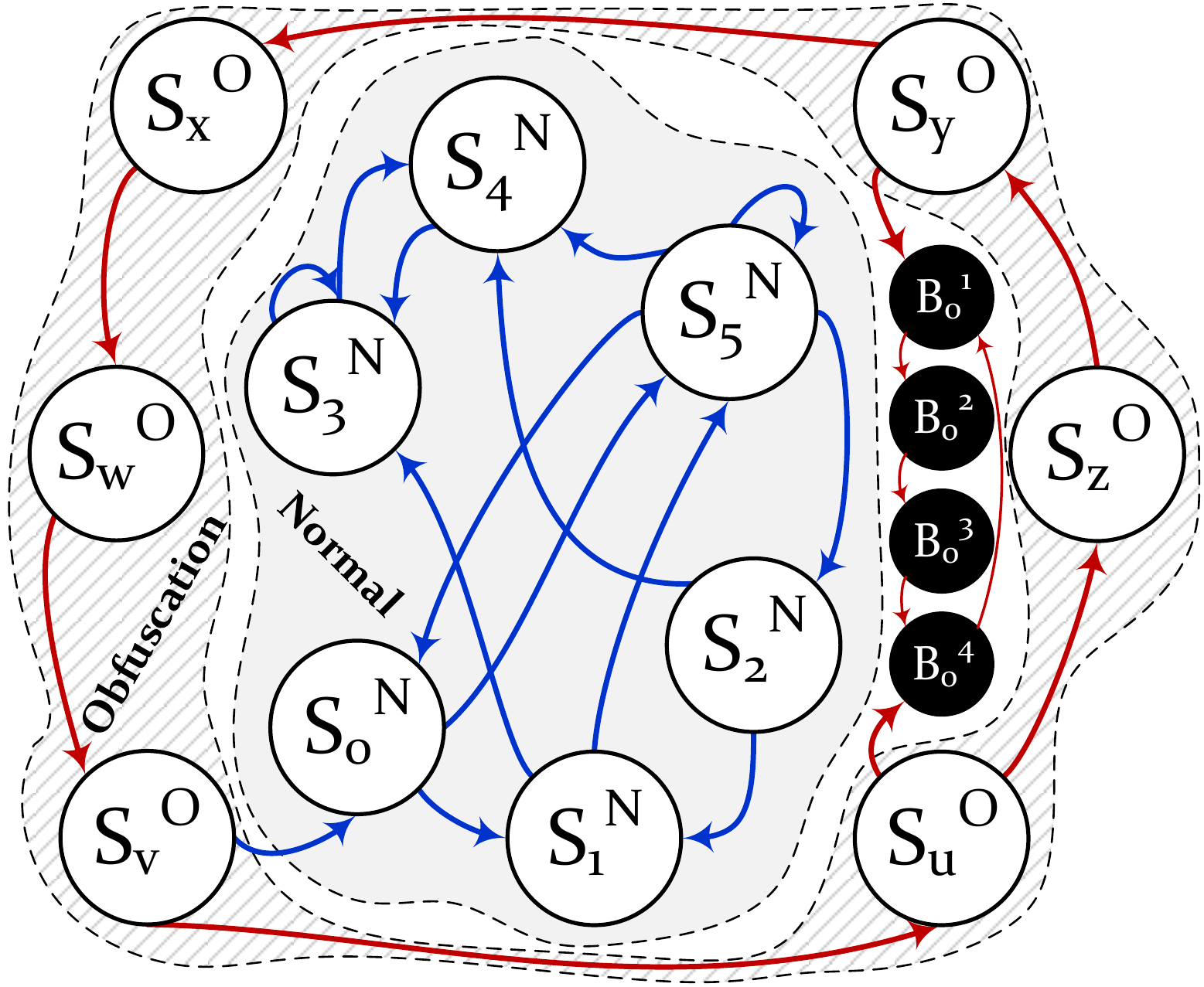} }}%
\subfloat[Interlocking Obfuscation \cite{desai2013interlocking}]{{\includegraphics[width=0.48\columnwidth]{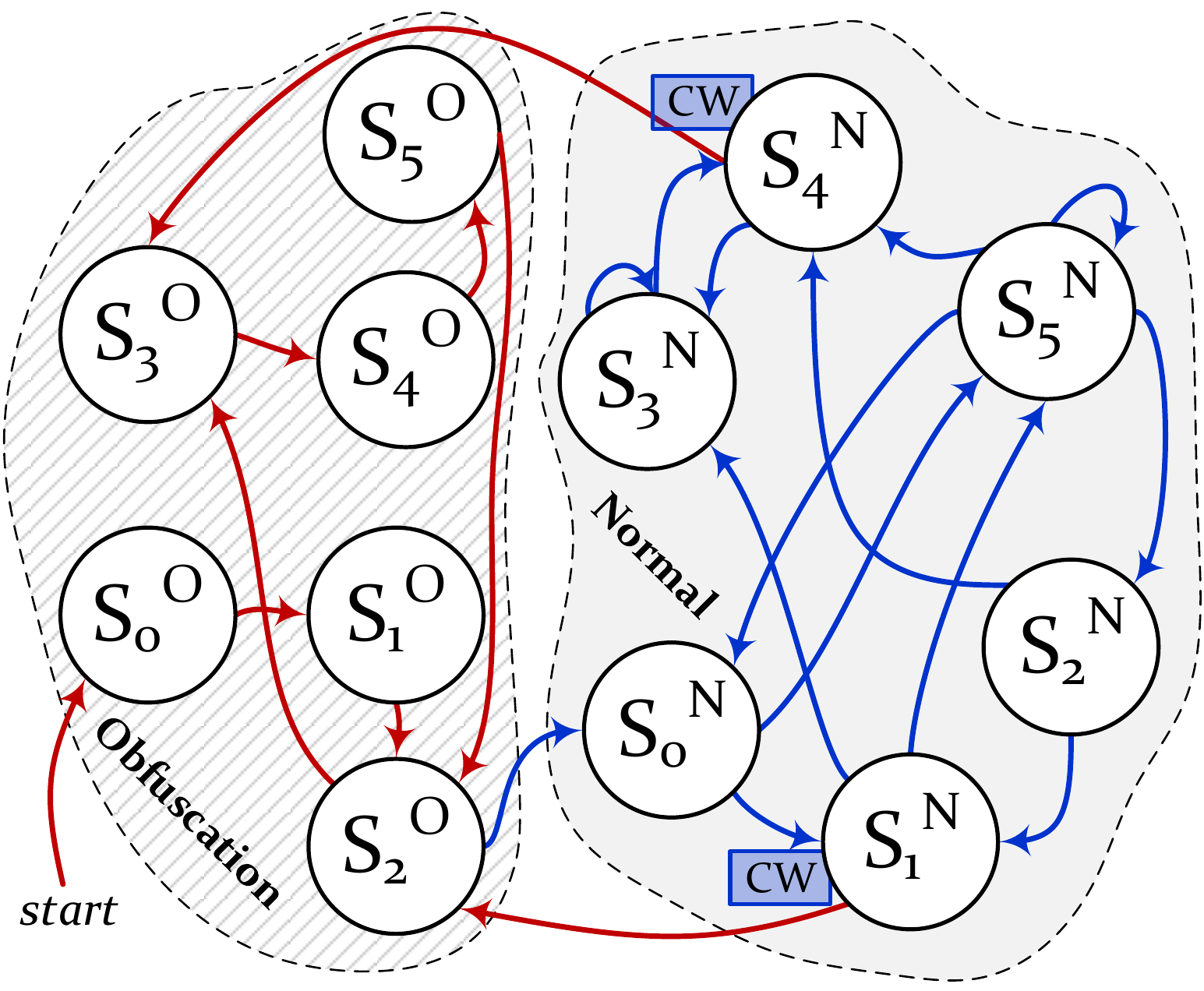} }}%
\caption{FSM Obfuscation Solutions.}
\label{scc}
\end{figure}

To assess the strength of FSM locking solutions, many studies evaluated the possibility of deploying 2-stage attacks, as an \emph{oracle-less} attack, on locked FSMs \cite{meade2017revisit, fyrbiak2018difficulty}. The 2-stage attacks on FSM are composed of: (1) \emph{\underline{\textbf{(stage 1): topological analysis}}} (described in line 2-13 of Algorithm \ref{topofunc}), which is a detection algorithm to find FFs that are responsible for storing the state values (separating them from datapath FFs), and \emph{\underline{\textbf{(stage 2): functional analysis}}} (described in line 14-21 of Algorithm \ref{topofunc}) that finds the STG based on the list of FFs found in (stage 1). In such an attack, the topological analysis, which is derived from \cite{shi2010highly}, identifies FFs whose input contains a combinational feedback path from their output. Then, it reduces the set of possible state FFs by (a) grouping the FFs controlled by the same set of signals, and (b) finding strongly connected components (SCC) using Tarjan's algorithm \cite{tarjan1972depth, meade2016netlist, meade2017revisit, fyrbiak2018difficulty}. In the functional analysis stage (stage 2), the attacker attempts to extract/re-draw the STG. This is done by first attempting to find the initial state, and then identifying the reachable states by creating a reduced binary decision diagram (BDD) or using a SAT solver. After re-drawing STG by using a 2-stage attack, in most FSM locking solutions, the adversary can readily distinguish the original part of the FSM from either extra added states or extra state transitions, leading to extracting the original FSM. Fig. \ref{scc} illustrates two examples of FSM locking. As shown, the original part of the FSM is easily distinguishable from extra locked states in these solutions. 

\begin{algorithm}[b]
\caption{2-stage on FSM Locking \cite{meade2017revisit, fyrbiak2018difficulty}}
\label{topofunc}
\begin{algorithmic}[1]
\footnotesize
\Function{FSM\_Extract}{Circuit~C$_{L}$}
    \State \emph{SFF} $\gets$ $[]$; \Comment{State Flip Flops}
    \State \emph{RS} $\gets$ classify(FFs); \Comment{Classifying FFs into Register Sets}
    \ForEach{\emph{set} $\in$ \emph{RS}}
        \State \emph{set} $\gets$ \emph{set} - {notSCC(\emph{set})}; \Comment{Keeps Strongly Connected Components}
        \If{is\_splittable(set)}
            \State \emph{RS} $\gets$ \{\emph{RS} - \emph{set}\} $\cup$ split(\emph{set});
        \EndIf
    \EndFor
    \State \emph{CLFP} $\gets$ find\_feedback\_circuits(C$_{L}$, Reg\_Sets);
    \ForEach{\emph{set} $\in$ \emph{RS}}
        \State \emph{set} $\gets$ \emph{set} - {notInfDep(\emph{set})}; \Comment{Keeps Intersected Influence/Dependence}
        \State \emph{set} $\gets$ \emph{set} - {InputIndependt(\emph{set})}; \Comment{Check Control Metrics}
        \State update(\emph{CLFP});
        \State SFF $\gets$ SFF $\cup$ \emph{set};
    \EndFor
    \State \emph{S$_0$} $\gets$ initial\_state(state\_regs); \emph{SQ} $\gets$ []; \Comment{State Queue}
    \State \emph{SQ} $\gets$ \emph{SQ} $\cup$ S$_0$; \emph{STF} $\gets$ []; \Comment{State Transition Table}
    \While{\emph{SQ} $\neq$ []}
        \State state $\gets$ \emph{SQ}.dequeue();
        \ForEach{\emph{DIP}} \Comment{DIP found by SAT}
            \If{eval(state\_regs, \emph{DIP}, state) $\notin$ \emph{SQ}}
                \State \emph{SQ}.enqueue(nx\_state);
                \State \emph{STF} $\gets$ \emph{STF} $\cup$ \{state, \emph{DIP}, nx\_state, \emph{PO}\}
            \EndIf
        \EndFor
    \EndWhile
    \Return \emph{SQ}, \emph{S$_0$}, \emph{STF}; \Comment{States, Initial, Transition Func.}
\EndFunction
\end{algorithmic}
\end{algorithm}

In UB-SAT or BMC \cite{el2017reverse, shamsi2019kc2, alrahis2019scansat} on the other hand, as a much stronger attack that is applicable to all FSM locking, sequential datapath locking, and scan chain locking, the adversary does not need to have access to the scan chain. In these attacks, the adversary unrolls the reverse-engineered netlist $n$ times and then creates a double circuit similar to the SAT attack. Then, the adversary uses a SAT solver to find a sequence of $n$ inputs and two key values such that the output of the meter (double) circuit detects a mismatch. Such an input sequence is denoted as a \emph{discriminating input sequence} (DIS). The attacker increases the unrolling depth ($n$) until a termination condition is met. The overall flow of this breed of attacks is captured in Algorithm \ref{scanfree}. By using this structure, the adversary can target and attack any sequential logic locking solution even while the access to the scan chain is restricted. Also, as an enhanced version of this group of attacks, KC2 \cite{shamsi2019kc2}, accelerates the original UB-SAT \cite{el2017reverse} by using some additional simplification steps. Some of the added features include (a) integrating BMC with SAT, (b) model conversion to BDD to simplify the circuit representation, and (c) constraint simplification using key-sweeping. Similarly, the ScanSAT \cite{alrahis2019scansat} is another unrolling-based SAT attack that only focuses on scan chain locking solutions, which is applicable to both statically and dynamically scan chain locking solutions.

\begin{algorithm}
\caption{UB-SAT on Sequential/Scan Locking \cite{el2017reverse, shamsi2019kc2, alrahis2019scansat}}
\label{scanfree}
\begin{algorithmic}[1]
\footnotesize
\Function{\emph{UB\_SAT}}{\emph{Circuit~C}}
\State \emph{b} $\gets$ \emph{initial\_boundary};
\State \emph{Terminated} $\gets$ \emph{False};
\State \emph{MC$_{circuit}$} $\gets$ C(X, K$_1$, Y$_1$) $\wedge$ C(X, K$_2$, Y$_2$) $\wedge$ (Y$_1$ $\ne$ Y$_2$);
\While {\emph{!Terminated}}
    \While {(X$_{DIS}$, K$_1$, K$_2$) $\leftarrow$ BMC(MC$_{circuit}$, b) = \emph{T}} 
        \State Y$_f$ $\leftarrow$ C$_{BlackBox}$(X$_{DI}$);
        \State MC$_{circuit}$ =$\wedge$ ~C(X$_{DIS}$, K$_1$, Y$_f$) $\wedge$ C(X$_{DIS}$, K$_2$, Y$_f$);
    \EndWhile

    \If{not BMC(MC$_{circuit}$, \emph{b})} \Comment{UC}
        \State \emph{Terminated} = \emph{True};
    \ElsIf{not BMC(MC$_{comb\_circuit}$, \emph{b})} \Comment{CE}
        \State \emph{Terminated} = \emph{True};
    \ElsIf{\emph{UMC}(MC$_{circuit}$, \emph{b})} \Comment{UMC}
        \State \emph{Terminated} = \emph{True};
    \Else
        \State \emph{b} $\gets$ \emph{b} + \emph{boundary\_step};
    \EndIf
\EndWhile
\State \emph{KeyGenCircuit} = DIVC $\wedge$ (K$_1$ = K$_2$)
\State \emph{Key} $\leftarrow$ BMC(emph{KeyGenCircuit}, \emph{b})
\EndFunction
\end{algorithmic}
\end{algorithm}

\section{Proposed Scheme: SCRAMBLE} \label{proposed}

\begin{figure}[b]
\centering
\subfloat[]{{\includegraphics[width=0.29\columnwidth]{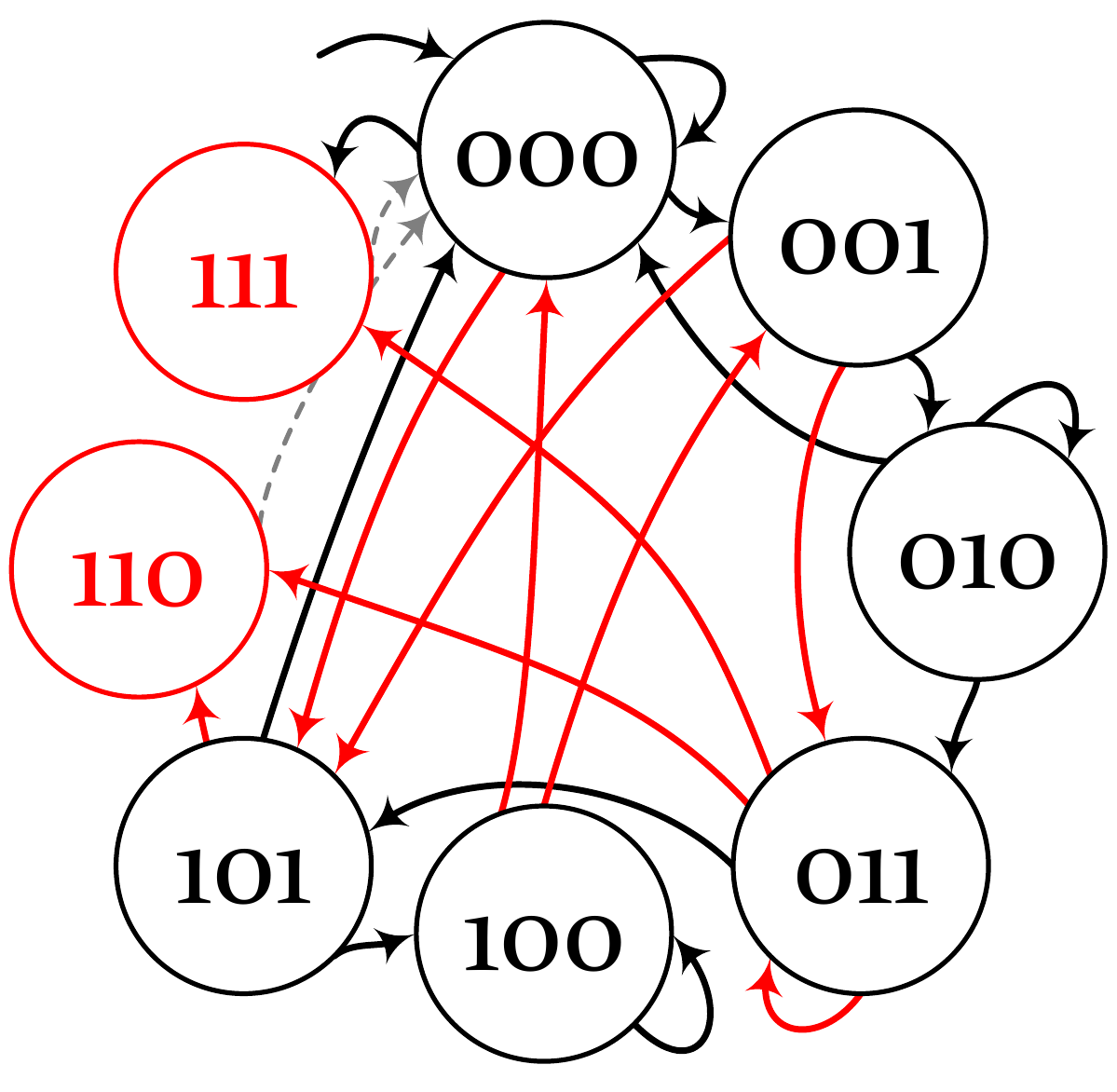} }}%
\subfloat[]{{\includegraphics[width=0.29\columnwidth]{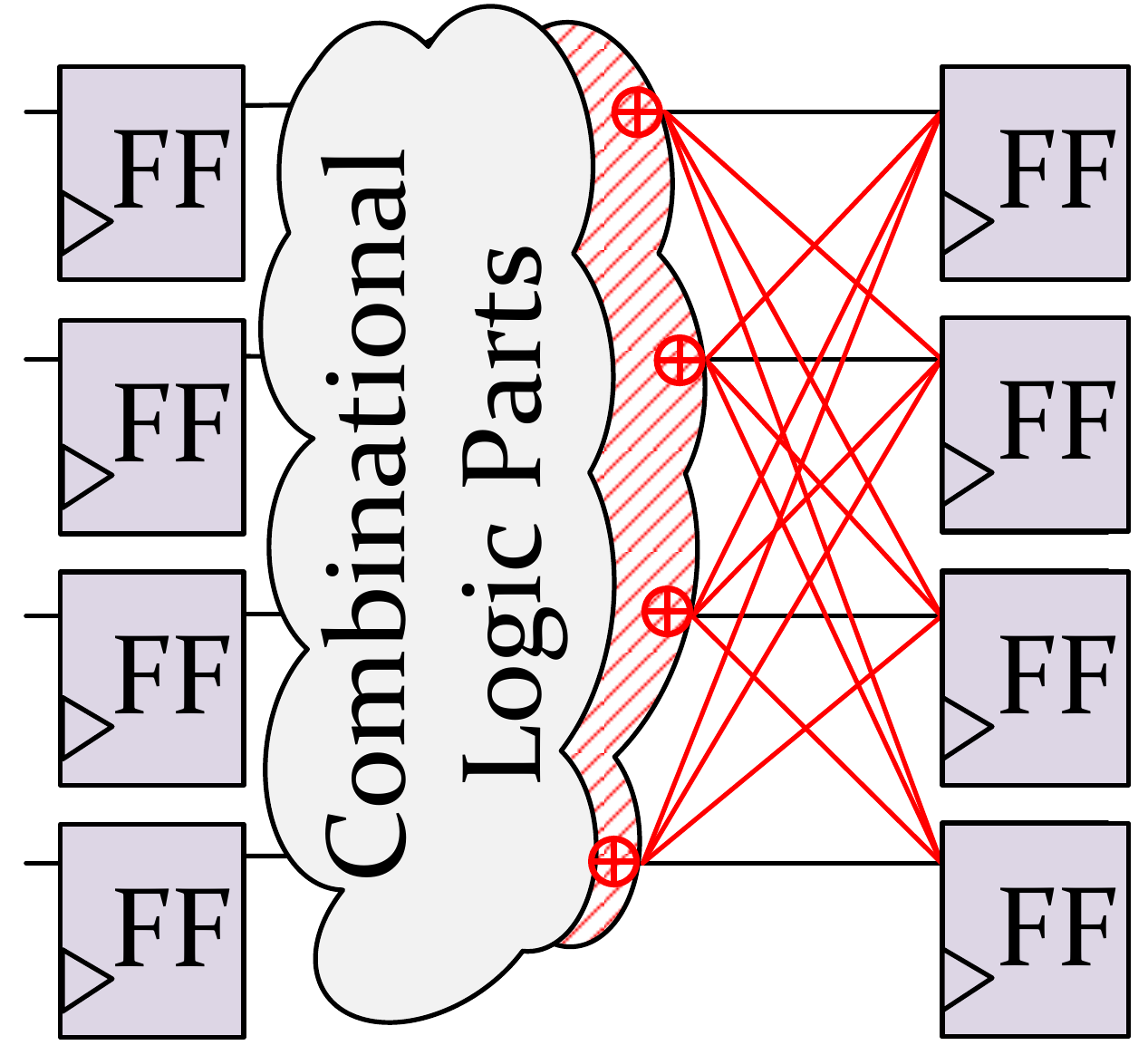} }}
\subfloat[]{{\includegraphics[width=0.38\columnwidth]{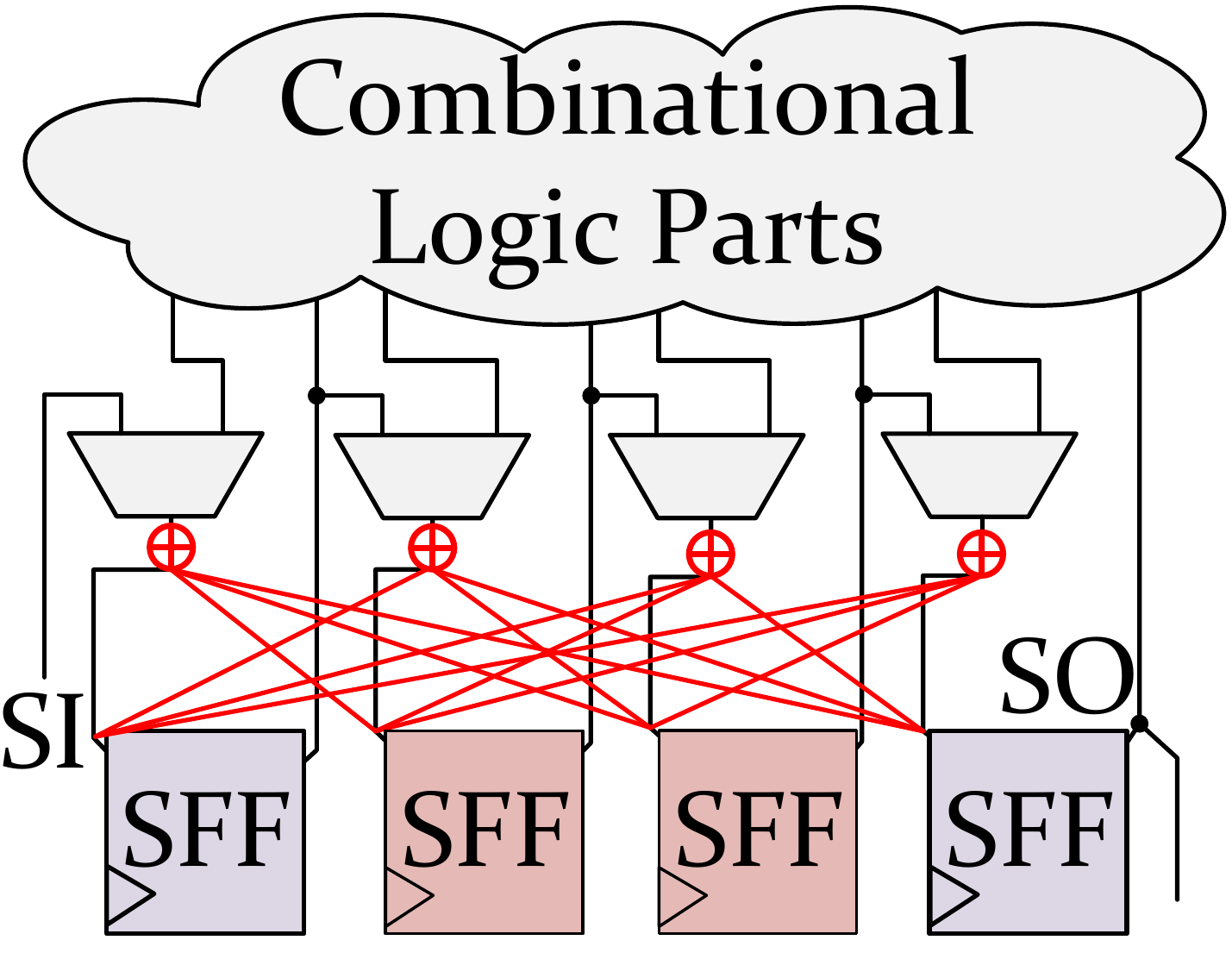} }}
\caption{\emph{Augmentation} for (a) FSM Locking, (b) Sequential Datapath Locking, and (c) Scan Chain Locking. (Black Lines: Original. Red lines: False by SCRAMBLE.)}
\label{define}
\end{figure}

In SCRAMBLE, we engage the term \emph{augmentation} to refer to the process illustrated in Fig. \ref{define}. \emph{Augmentation} in SCRAMBLE adds false state transitions in case of FSM locking, or adds false FF-to-FF timing paths in case of sequential datapath locking, or adds false scan chain sequence in case of scan chain locking. SCRAMBLE is proposed in two variants: (1) The first variant is \emph{connectivity} SCRAMBLE (SCRAMBLE-C) that hides the connectivity to the targeted FFs using logarithmic switching network. (2) The second variant is \emph{logic} SCRAMBLE (SCRAMBLE-L) that hides the logic by implementing part(s) of the logic within memory. The SCRAMBLE-C could be used for locking either FSMs, sequential datapath, or scan chains, to protect the locked design against all UB-SAT and BMC attacks, such as KC2 or ScanSAT. SCRAMBLE-L, on the other hand, is mostly applicable to FSMs to provide resilience against 2-stage attacks \cite{meade2017revisit, fyrbiak2018difficulty}.

\subsection{SCRAMBLE-C}\label{section-scramble-n}

\begin{figure}[t]
\centering
\includegraphics[width = 240pt]{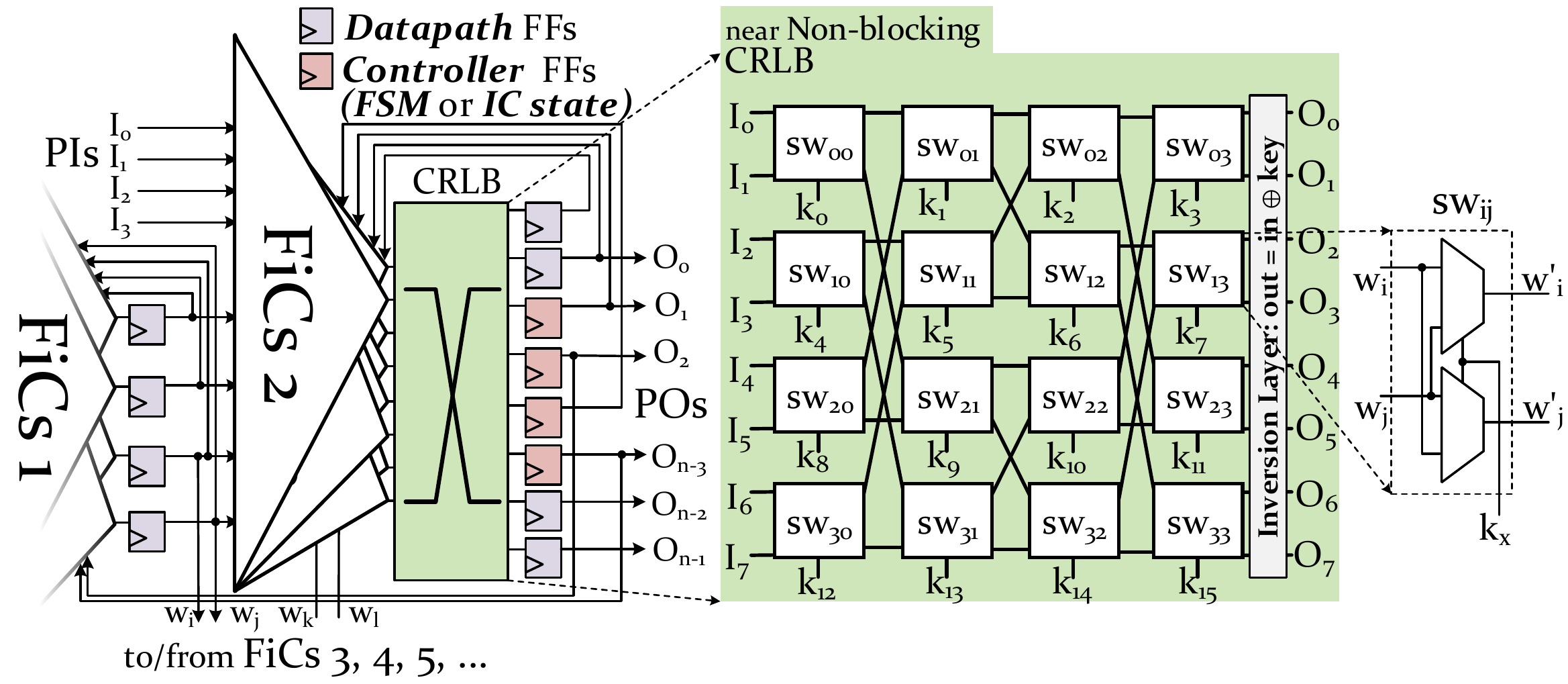}
\caption{\emph{Augmentation} via Inserting CRLB (\emph{shuffle}-based) in SCRAMBLE-C.}
\label{fig-scramble-c}
\end{figure}

The overall structure of SCRAMBLE-C has been illustrated in Fig. \ref{fig-scramble-c}. In SCRAMBLE-C, the connectivity between the targeted FFs and their \emph{fan-in-cone}s (FiCs) (combinational logic cones) is locked. Hence, before connecting the output of corresponding FiCs to the FFs, a \emph{configurable routing and logic block} (CRLB) has been inserted to control the connections. For instance, in Fig. \ref{fig-scramble-c}, a CRLB with size 8 has been inserted before a combination of FSM FFs and datapath FFs. The CRLB, which must be inserted between the targeted FFs and their corresponding FiCs, is a key-programmable switching network that is built using self-routing logarithmic ($log_{2}(N)$) networks \cite{shyy1991log, kamali2019full}. The $log_{2}(N)$ networks, compared to the existing and well-known switching networks, such as mesh, crossbar, etc., incur less area overhead. Also, we demonstrate that due to its cascading structure, $log_{2}(N)$ networks could improve the robustness against the SAT attack.

As shown in Fig. \ref{fig-scramble-c}, the CRLB is built using key-programmable 2x2 switch-boxes ($sw_{ij}$). Based on its key, a $sw$ saves or changes the order of inputs while connecting them to output pins. Also, the connection between the layers of $sw$s is fixed. This inter-layer connection determines the topology of the $log_{2}(N)$ network. For instance, the architecture of a sample CRLB (\emph{shuffle} topology) with size 8 has been demonstrated in Fig. \ref{fig-scramble-c}. Also, as shown in this example, to add the capability of logic programmability, we add one extra key-controlled (XOR) inversion layer, as the final layer of CRLB, to twist routing locking with logic locking. The addition of the inversion layer allows the CRLB to not only permute the inputs, but it also negates them based on the keys of the inversion layer.

In SCRAMBLE-C, the CRLB must be inserted before the targeted FFs. When FSM locking or sequential datapath locking is targeted, during either the physical design or after DFT synthesis step, the CRLB is placed on wires that connect the outputs of FiCs to the \emph{data-in (DI) pin} of targeted FSM FFs or datapath FFs. When scan chain locking is targeted, after DFT synthesis, the CRLB is placed in \emph{scan network} before the \emph{scan-in (SI) pin} of the targeted SFFs.

Although engaging self-routing $log_{2}(N)$ networks provides a low-overhead routing locking solution, we have to address a few issues: (1) The size of the $log_{2}(N)$ circuit grows exponentially as the input size grows. (2) The nature of $log_{2}(N)$ networks is blocking, and many of the input permutations cannot be routed. Hence, the number of false transitions/connections/sequences in FSM/datapath/scan would be very limited. Hence, the wires ($N$) as the inputs of CRLB must be small enough to make the network overhead reasonable; and large enough to make it resistant against SAT-driven attacks, i.e. UB-SAT or BMC. It raises two questions: (1) which $N$ FFs must be selected? and (2) How we can minimize $N$?  

\subsubsection{\textbf{Selection of $N$ FFs}}

\begin{figure}[t]
\centering
\subfloat[]{{\includegraphics[width=0.98\columnwidth]{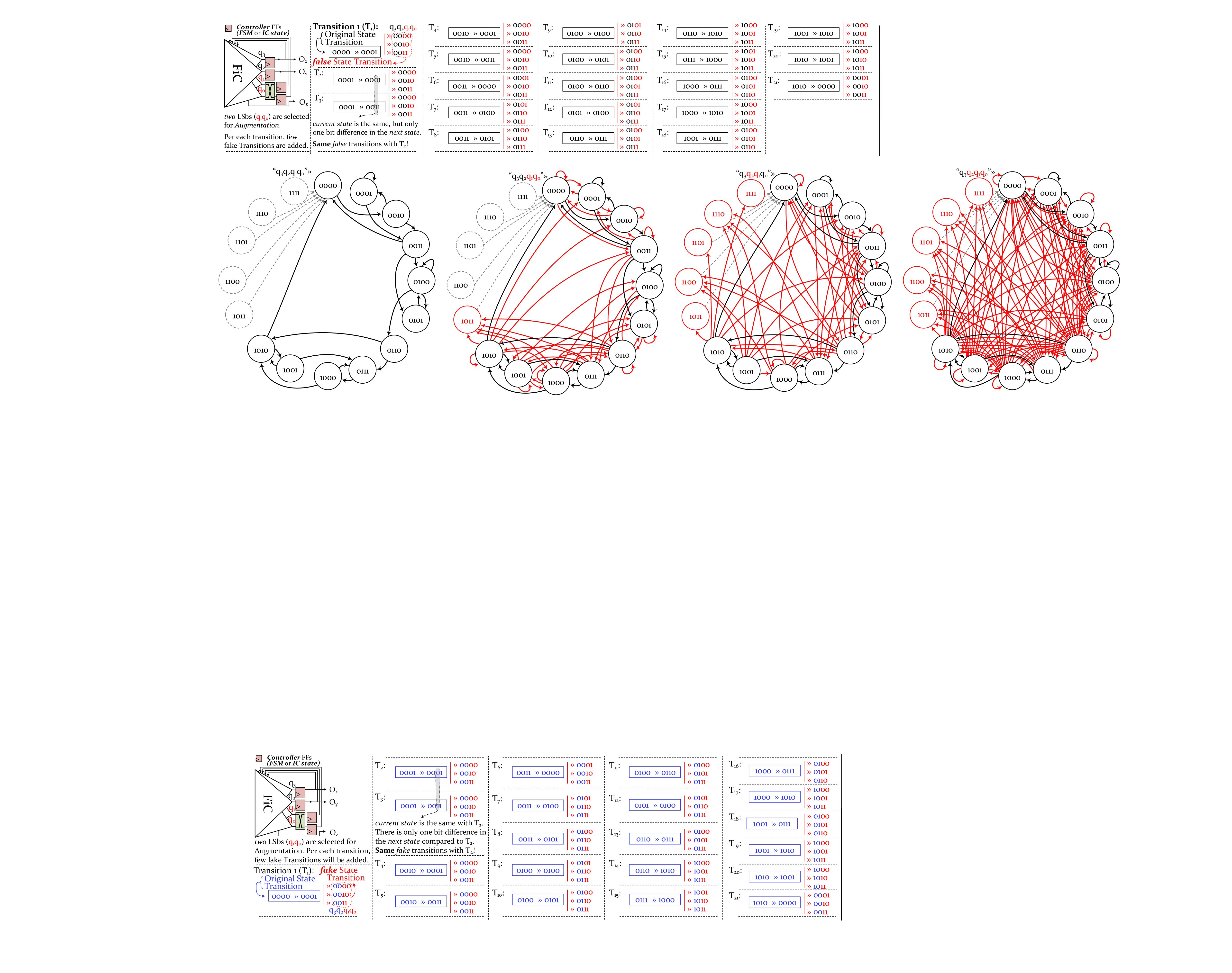} }} \\ 
\subfloat[]{{\includegraphics[width=0.47\columnwidth]{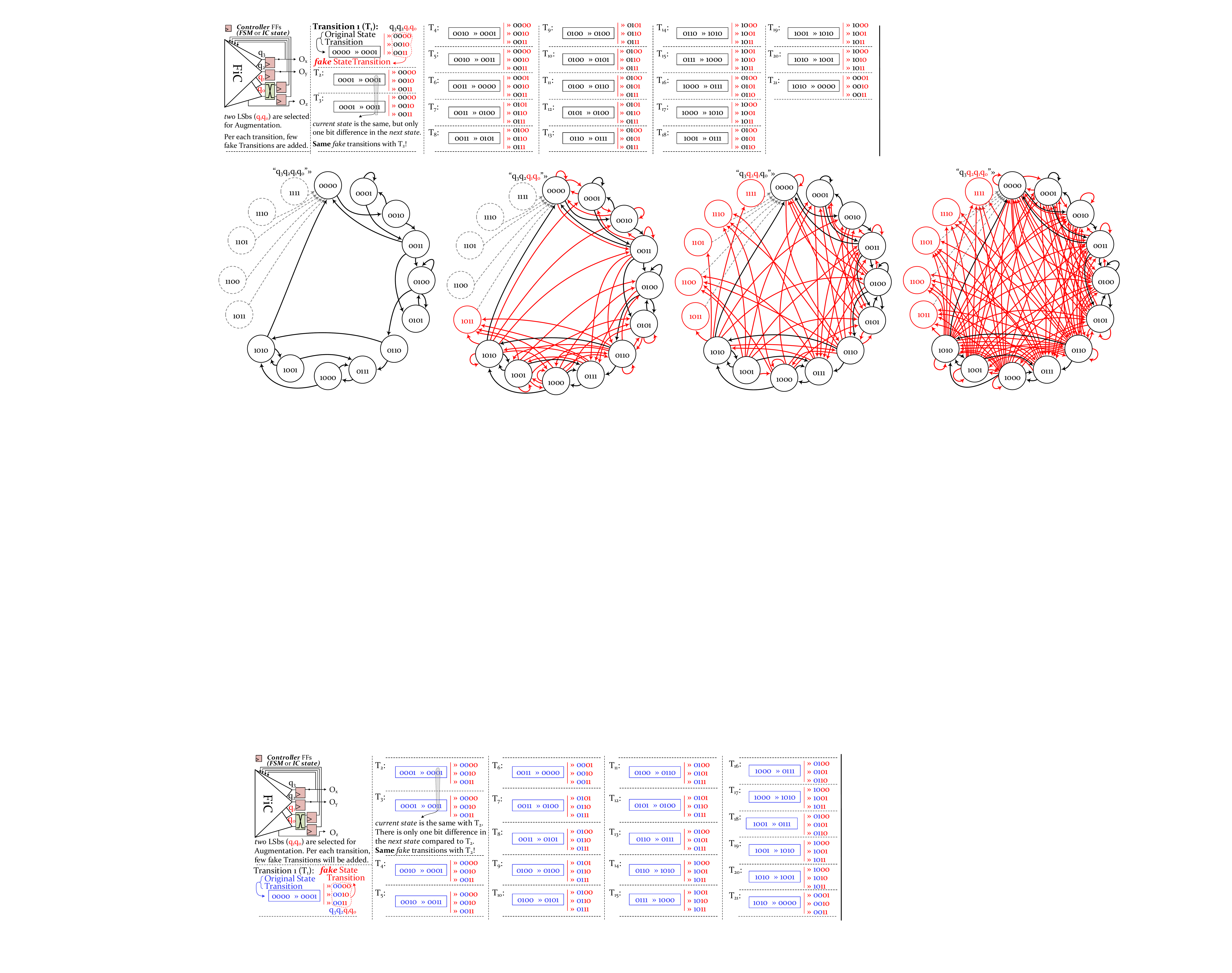} }} 
\subfloat[]{{\includegraphics[width=0.47\columnwidth]{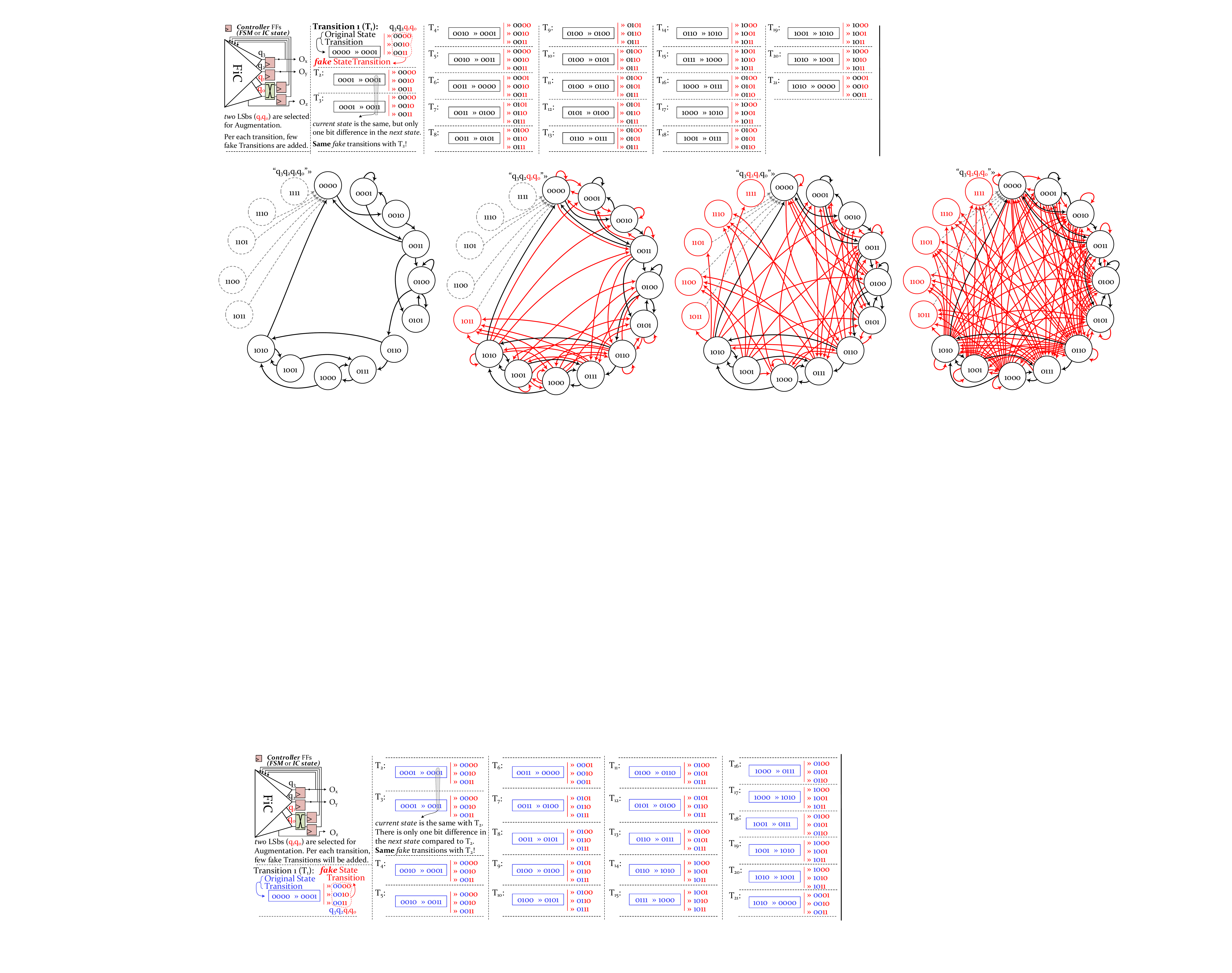} }} \\ 
\subfloat[]{{\includegraphics[width=0.47\columnwidth]{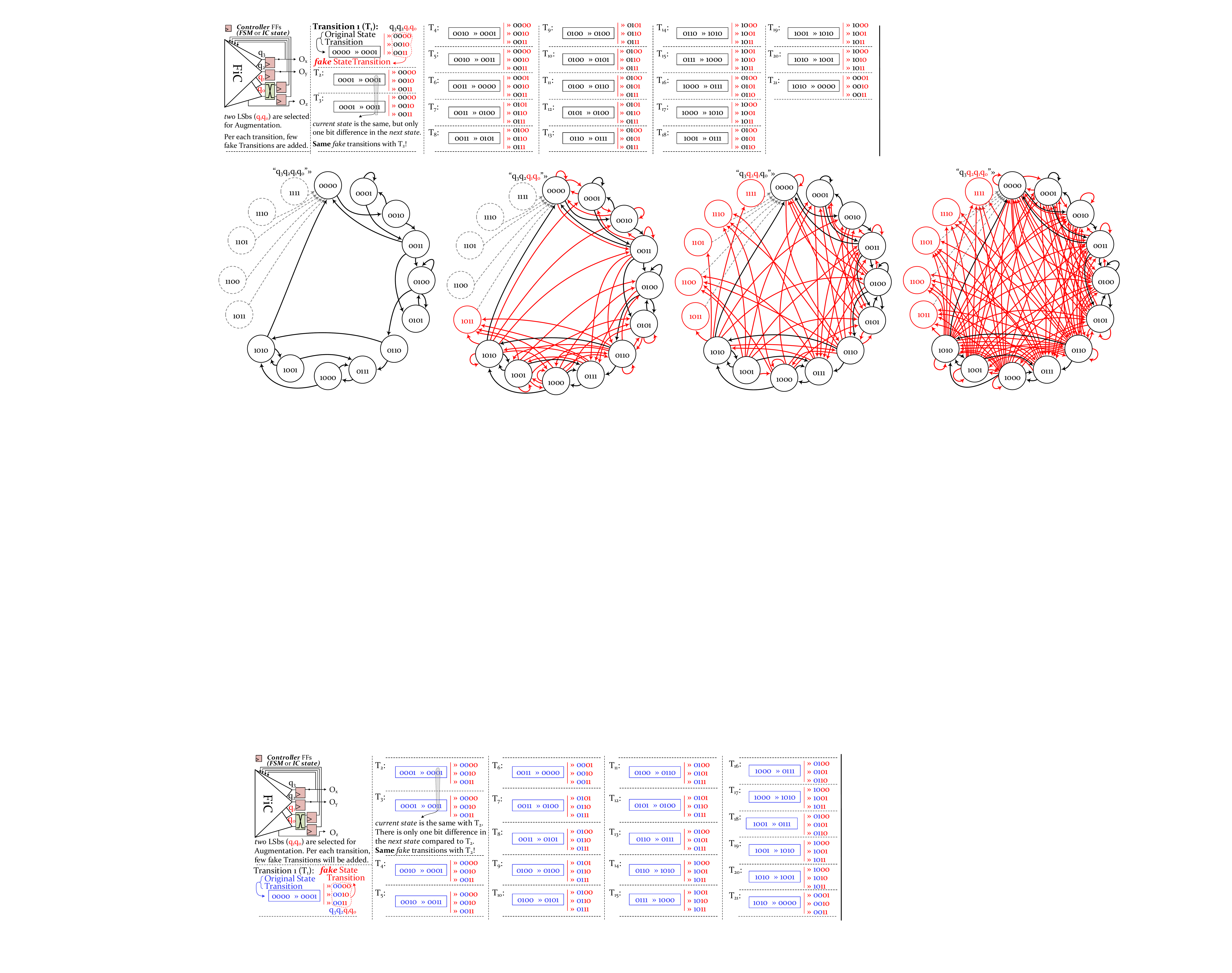} }}
\subfloat[]{{\includegraphics[width=0.47\columnwidth]{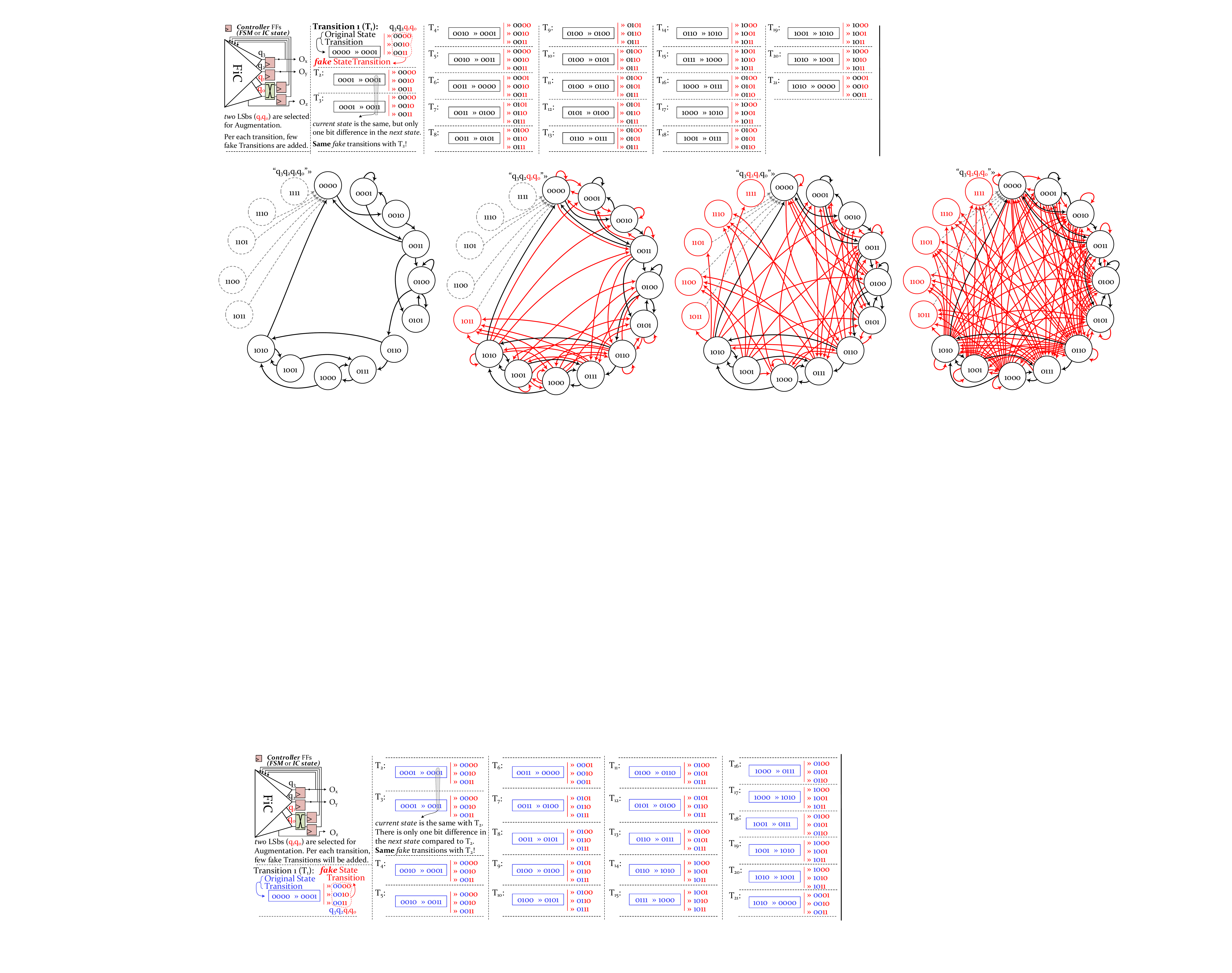} }}
\caption{Using SCRAMBLE-C for FSM Locking (a) FSM circuit and transitions generation, (b) the original FSM, (c) 2 LSBs are inputs to SCRAMBLE-C, (d) 2 MSBs are inputs to SCRAMBLE-C, (e) 3 LSBs are inputs to SCRAMBLE-C.}
\label{fig-scramble-p}
\end{figure}

The selection of FFs (to insert CRLB before them) in SCRAMBLE-C could significantly impact its locking strength, particularly in FSM locking. For example, let us consider the engaging of SCRAMBLE-C for an FSM presented in Fig. \ref{fig-scramble-p}(b), which is generated using Binary encoding of 4 FFs. In this example, if we select \emph{two} least significant bits (LSB) FFs to insert a CRLB with size 2 before them (Circuit of Fig. \ref{fig-scramble-p}(a)), the locked FSM is demonstrated in Fig. \ref{fig-scramble-p}(c). Fig. \ref{fig-scramble-p}(a) shows how the false transitions in Fig. \ref{fig-scramble-p}(c) have been generated for some of the original transitions. As shown in \ref{fig-scramble-p}(c), when 2 LSBs are selected, a limited number of false transitions are added, and only one extra (unreachable) state is visited. However, if we select \emph{two} most significant bits (MSB) FFs, it will visit all extra states and generates a large number of false transitions demonstrated in Fig. \ref{fig-scramble-p}(d). Since 2-stage attacks are applicable to FSM locking, maximizing false transitions as well as extra states makes SCRAMBLE-C more robust against this attack. Accordingly, being aware of the encoding style of FSM will impact its locking strength. For instance, in Binary encoding, a synthesis tool usually encodes the states from low to high (0 to $2^{N-1}$). Hence, using the $N$ FFs representing the MSB of state value results in the inclusion of the largest number of previously unreachable states and false transitions in the locked FSM. Also, Fig. \ref{fig-scramble-p}(e) shows locked FSM when \emph{three} LSB FFs are selected. It shows that even increasing the size of CRLB by one adds numerous false state transitions into the locked FSM. Note that unlike FSM locking, the selection of $N$ FFs has no impact on locking strength when sequential datapath locking or scan chain locking is targeted.

\subsubsection{\textbf{Reducing $N$ by making CRLBs near non-blocking}}

The implementation of blocking $log_{2}(N)$ network revealed that even a 256-input CRLB could be broken by SAT attack in less than an hour. Hence, to address the blocking nature of CRLB and to resist against UB-SAT or BMC attacks (with a small CRLB), we expand the $log_{2}(N)$ network towards non-blocking via adding extra (cascaded) stages. The expanded $log_{2}(N)$ network with strictly non-blocking structure is generalized under the notation $LOG_{2}(N, M, P)$, where $N$ denotes the number of inputs, $M$ is the number of extra (cascaded) stages, and $P$ indicates that there are $P-1$ additional copies (of whole network) \emph{vertically cascaded} \cite{shyy1991log}. However, For a network with an arbitrary $N$, the minimum value of $M$ and $P$ to make the network strictly non-blocking are extremely large. For instance, with $N=64$ the choice of $M$ and $P$ should be 3 and 6 respectively, resulting in 5x area overhead compared to a blocking $log_{2}(N)$ \cite{shyy1991log}. Hence, strictly non-blocking incurs almost prohibited area overhead.

To move close enough towards non-blocking nature without incurring large area overhead, we used the \emph{"near non-blocking"} structure \cite{shyy1991log}. In near non-blocking, not all but \emph{almost all} permutations are feasible, while it could be implemented using a $LOG_{2}(N, log_2(N) - 2, 1)$, meaning it has only $M = log_2(N) - 2$ extra stages and no additional copy ($P = 1$). The CRLB depicted in Fig. \ref{fig-scramble-c} is an example of a near non-blocking CRLB for 8 inputs. Our implementation shows that a 32-input near non-blocking network ($LOG_{2}(32, 3, 1)$) is far stronger against SAT attack compared to a 256-input blocking network $log_{2}(256)$, while it is 8x smaller. 

\subsection{SCRAMBLE-L} \label{section-SCRAMBLE-L}

\begin{figure}[t]
\centering
\subfloat[]{{\includegraphics[width=0.45\columnwidth]{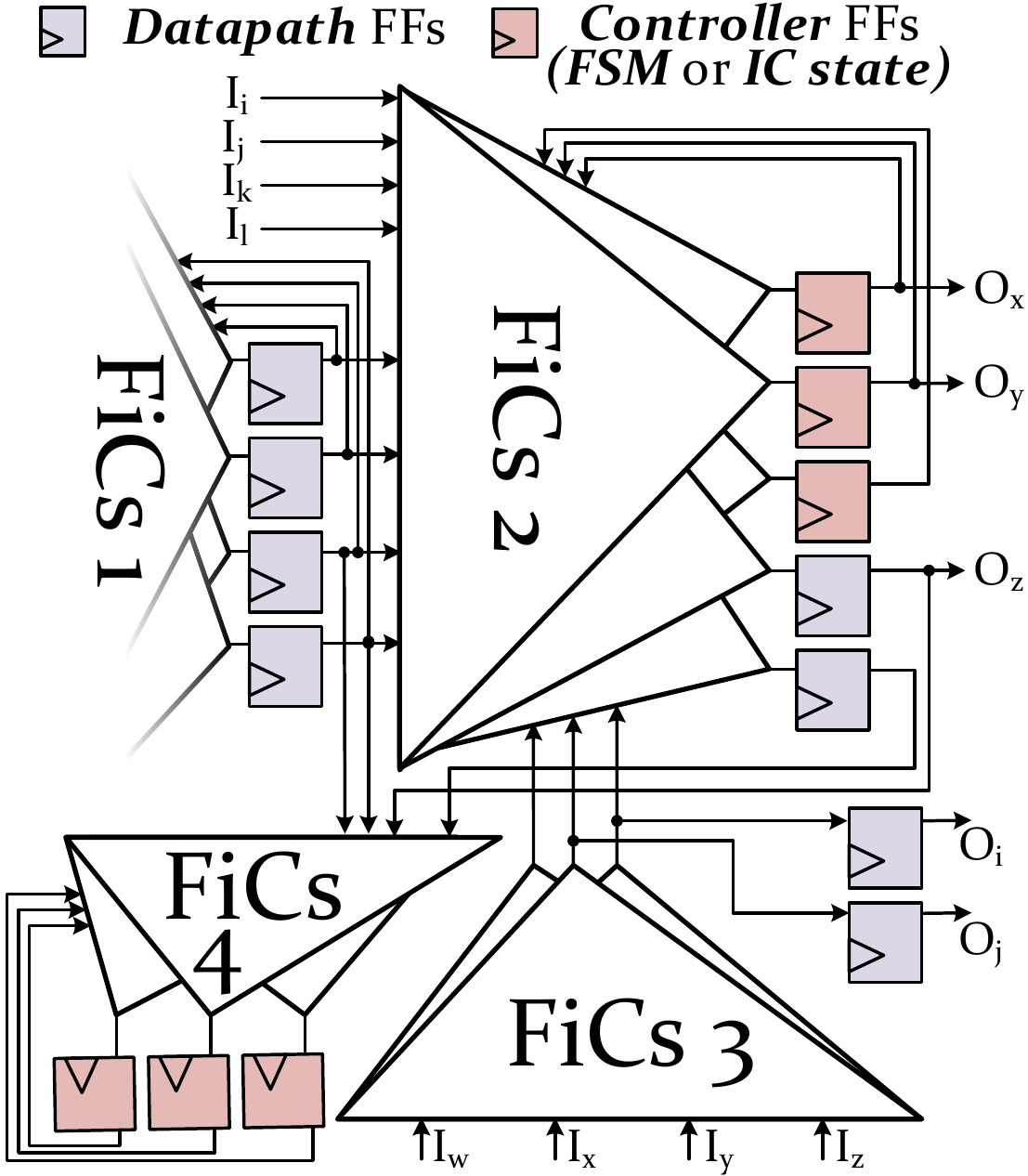} }}%
\subfloat[]{{\includegraphics[width=0.45\columnwidth]{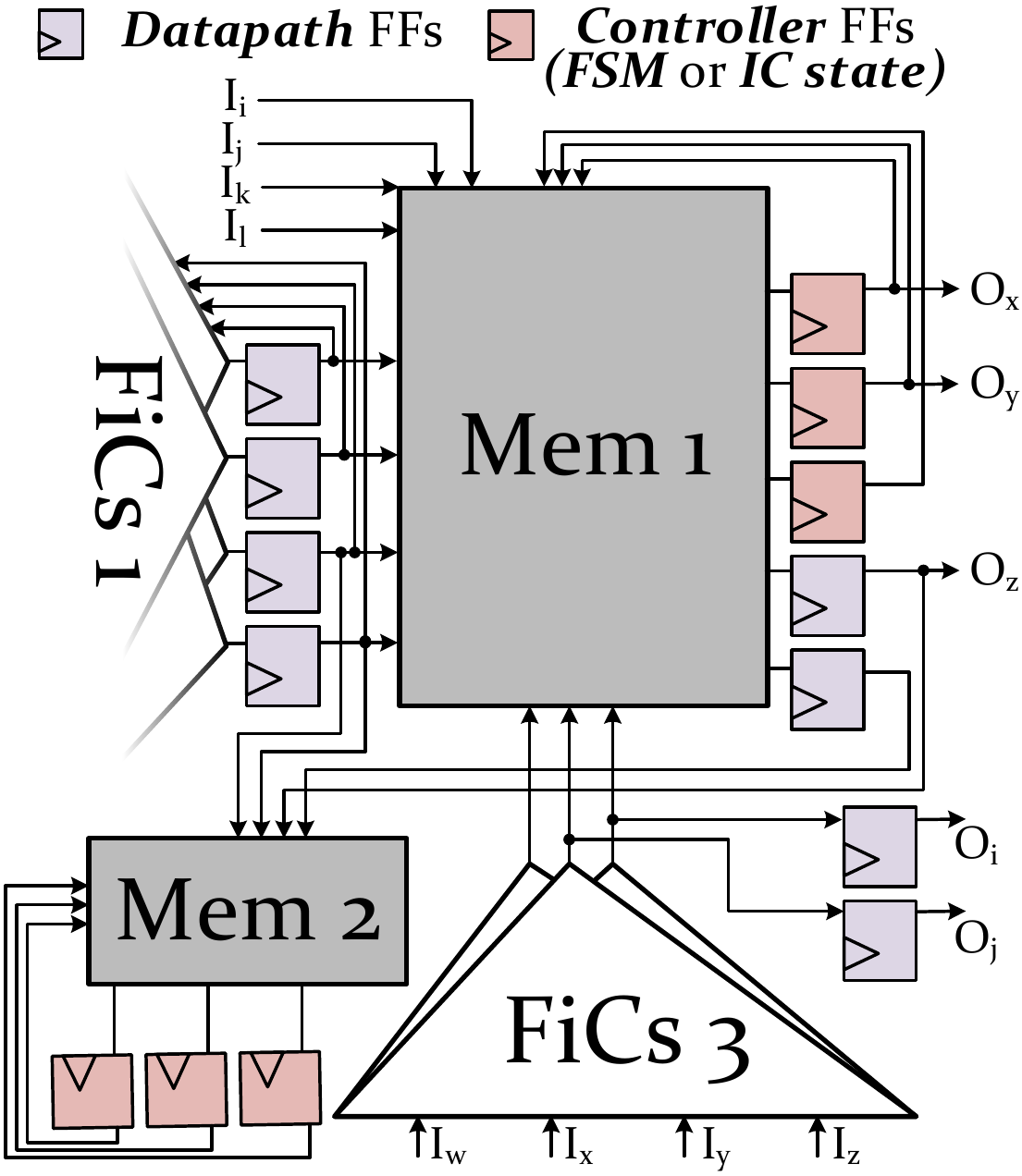} }}
\caption{Sequential Circuits Locking using SCRAMBLE-L.}
\label{fig-scramble-l}
\end{figure}

In SCRAMBLE-L, which is proposed for FSM locking against 2-stage attacks, the logic before the targeted FFs is locked using in-memory computation. As shown in Fig. \ref{fig-scramble-l}, a small part of the combinational logic in the FiCs of the targeted FFs is replaced with a one-cycle read memory, such as SRAM. As an example, $FiC_2$ and $FiC_4$ are replaced with equivalent memories. The content of the memories must provide the same output compared to that of $FiC_2$ and $FiC_4$ while the triggering input is the same. Hence, the truth table corresponded to those FiCs must be generated and stored in the memories. The memories would be initialized during boot-up from a tamper-proof NVM which serves as the key storage. To avoid additional delay incurred by memories, the selection of FFs must be done with respect to their available timing slack. 

SCRAMBLE-L makes 2-stage attacks almost impractical. Considering that the adversary has no access to the contents of memories after reverse engineering, there is no equivalent logic for the memories, and the BDD- or SAT-based \emph{functional analysis} (stage 2) cannot be accomplished on the locked circuit. Also, similar to Fig. \ref{fig-scramble-l}(b), if the designer selects a combination of datapath FFs and FSM FFs, the adversary cannot distinguish between them when deploying \emph{topological analysis} (stage 1) of the 2-stage attack, resulting in the inclusion of an extremely large number of non-FSM FFs in the candidate FSM FFs. Hence, none of the existing 2-stage attacks can be applied to SCRAMBLE-L.

\begin{figure}[t]
\centering
\subfloat[]{{\includegraphics[width=0.24\columnwidth]{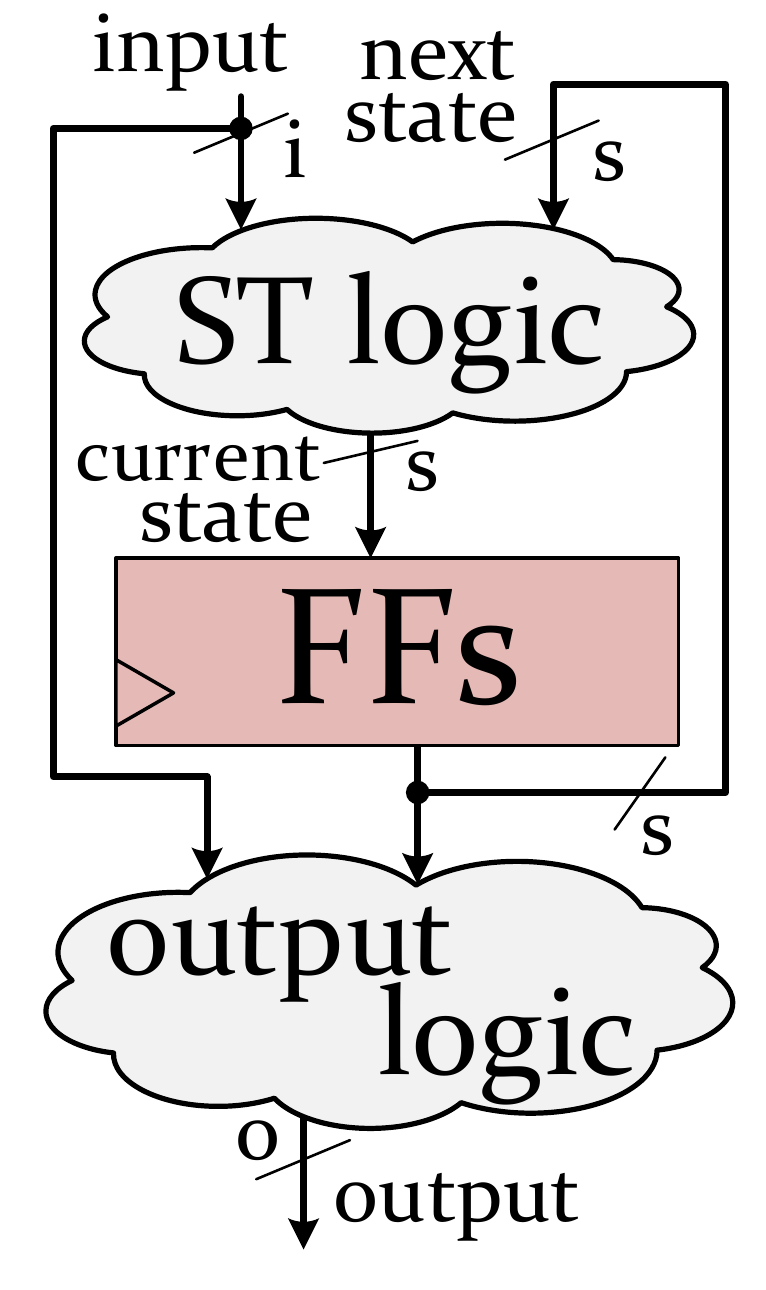} }}%
\subfloat[]{{\includegraphics[width=0.24\columnwidth]{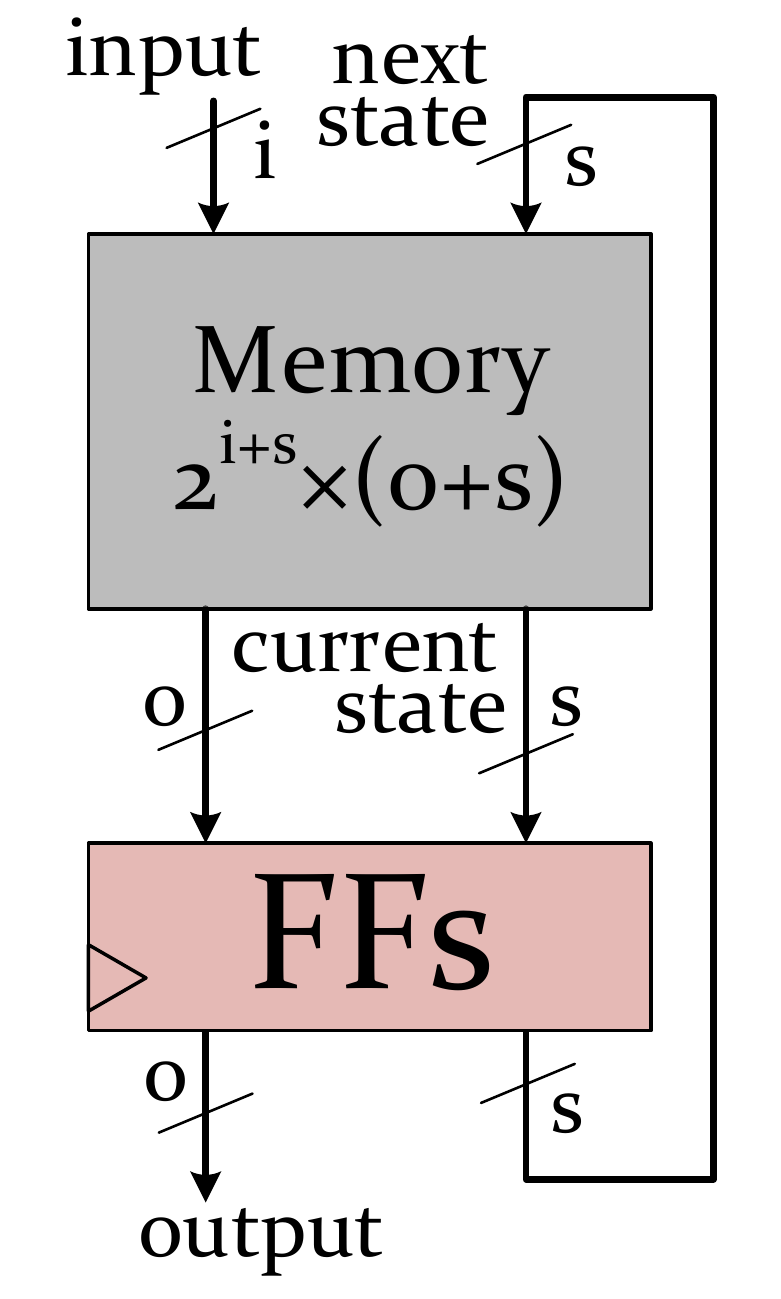} }}%
\subfloat[]{{\includegraphics[width=0.24\columnwidth]{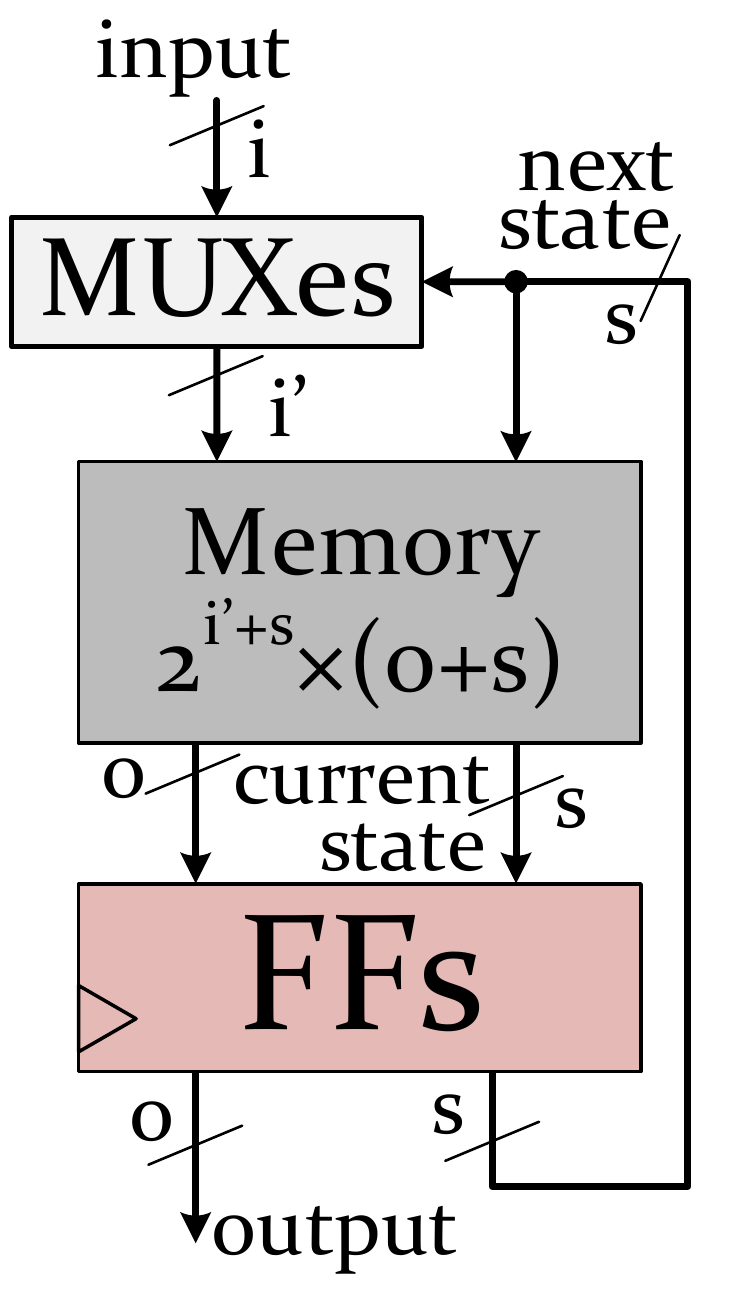} }}%
\subfloat[]{{\includegraphics[width=0.24\columnwidth]{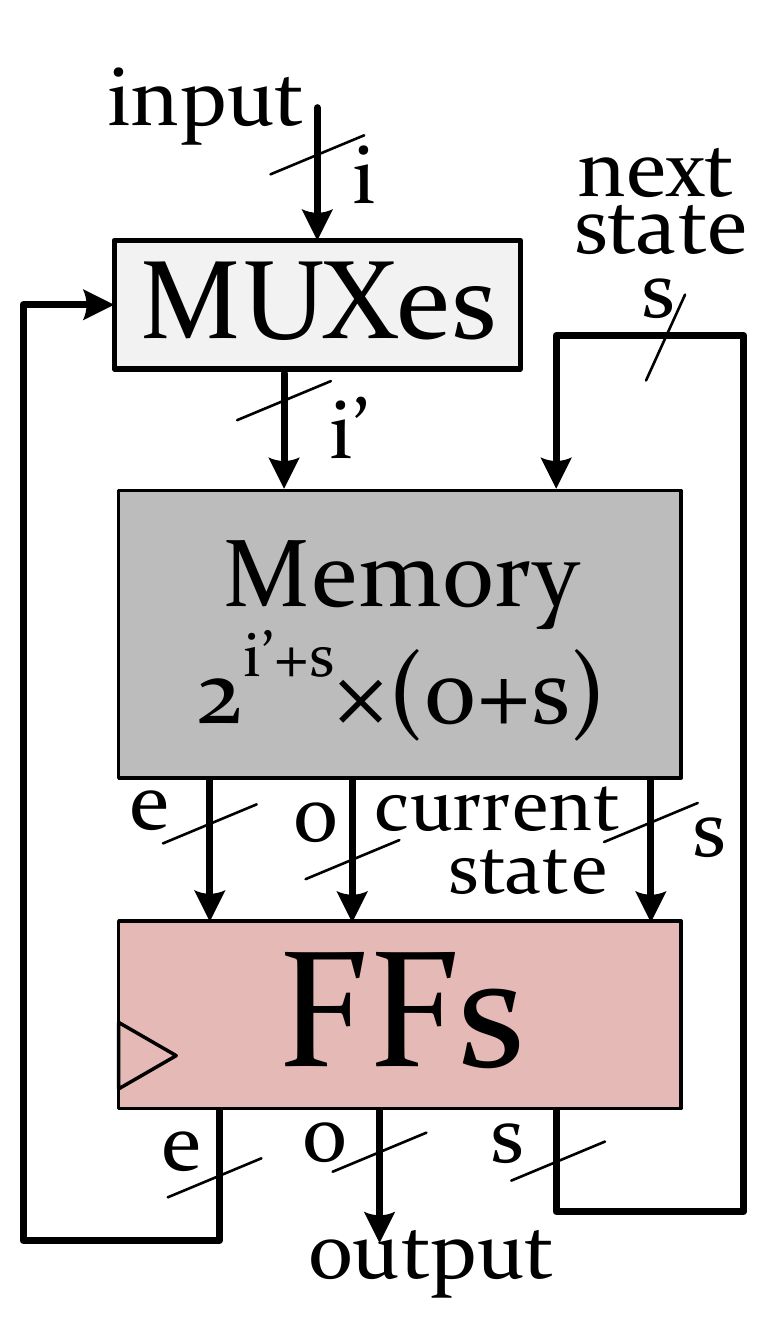} }} 
\caption{FSM Implementation (a) Traditional, (b) Memory-based, (c) Memory-based with Input Multiplexing using Current State, (d) Memory-based with Input Multiplexing using Code-word stored in Memory.}
\label{inputmux}
\end{figure}

The big challenge with the SCRAMBLE-L is the size of the memory for implementing the selected FiCs. However, since SCRAMBLE-L is proposed for FSM locking, this problem could be easily addressed by engaging the FSM input multiplexing (FSMIM) techniques \cite{garcia2007rom}. In this technique, considering that the next state and the outputs of the FSM are a function of a subset of the inputs (not all), a set of multiplexers has been used to select only those signals that affect the next state and the output. Hence, the designer is able to minimize the number of inputs to the memories (as address width), resulting in a significant decrease in the size of memory. The main difference between traditional FSM implementation, memory-based FSM, and FSMIM has shown in Fig. \ref{inputmux}.

In FSMIM, multiplexers could be controlled using two different strategies: (1) using the value of the current state, (2) using code-words stored in the memory. The first option is more efficient in terms of memory size reduction; however, the second method has better efficiency in reducing the multiplexers complexity. Hence, the first option has been used in SCRAMBLE-L to minimize the area overhead of the memories. Our evaluation in Table \ref{fsmreduce} illustrates the effectiveness of FSMIM when applied to the ISCAS-89 benchmarks, resulting in memory size reduction above 90\%.    

\begin{table}[t]
\scriptsize 
\centering
\caption{Simplification Ratio of Input Multiplexing (FSMIM).}
\label{fsmreduce}
\setlength\tabcolsep{1pt} 
\begin{tabular}{@{} *{12}c @{}}
\toprule 
\multirow{3}{*}{FSM} &  & \multicolumn{7}{c}{Required Memory Size and Additional MUXes}  \\
        \cmidrule(lr){3-9}
& &  Traditional & \multicolumn{3}{c}{Input MUXing} & \multicolumn{3}{c}{Input MUXing \emph{+} State Reduce} \\
  \cmidrule(lr){3-3}
 \cmidrule(lr){4-6}
 \cmidrule(lr){7-9}
& & Size$_{Kb}$ & Size$_{Kb}$ & MUX & Reduction & Size$_{Kb}$ & MUX & Reduction \\
\cmidrule(r){1-2}
\cmidrule(lr){2-3}
\cmidrule(lr){4-6}
\cmidrule(lr){7-9}
s510 & & 435,500 & 5.5 & 14, 5 & $\sim$99.9\% & 2.5 & 14, 7 & $\sim$99.9\% \\
\cmidrule(r){1-2}
\cmidrule(lr){2-3}
\cmidrule(lr){4-6}
\cmidrule(lr){7-9}
s820 & & 195,000 & 255 & 5, 4, 3, 2 & $\sim$99.9\% & 38 & 7, 6, 4, 4, 2, 2. 2 & $\sim$99.9\% \\
\cmidrule(r){1-2}
\cmidrule(lr){2-3}
\cmidrule(lr){4-6}
\cmidrule(lr){7-9}
s832 & & 200,000 & 262.5 & 5, 4, 3, 2 & $\sim$99.9\% & 69 & 5, 4, 4, 4, 2 & $\sim$99.9\% \\
\cmidrule(r){1-2}
\cmidrule(lr){2-3}
\cmidrule(lr){4-6}
\cmidrule(lr){7-9}
s1488 & & 408,000 & 110,500 & 2, 2 & 73\% & 16,000 & 4, 4, 2, 2, 2 & 92.5\% \\
\cmidrule(r){1-2}
\cmidrule(lr){2-3}
\cmidrule(lr){4-6}
\cmidrule(lr){7-9}
s1494 & & 408,000 & 110,500 & 2, 2 & 73\% & 16,000 & 4, 4, 2, 2, 2 & 92.5\% \\
\bottomrule
\end{tabular}
\end{table}

\section{Discussion}

Table \ref{compare} shows the effectiveness of each variant of SCRAMBLE against 2-stage and UB-SAT or BMC attacks. Although the main aim of SCRAMBLE-C is to protect the design against UB-SAT and BMC, it also breaks 2-stage attacks. Similar to SCRAMBLE-L, if we use a combination of both datapath FFs and FSM FFs as input to SCRAMBLE-C (Similar to Fig. \ref{fig-scramble-c}), topological analysis (stage 1) of 2-stage attack cannot detect the correct set of FSM FFs. Therefore, the functional analysis (stage 2) has to generate the STG using an incorrect set of FFs (extremely larger set), resulting in a significant increase in the attack time with respect to the number of additional (datapath) FFs included in the set. Also, the extracted STG is constructed using a combination of datapath FFs and FSM FFs, which leads to an incorrect STG, and the adversary is not able to extract the original FSM from the extracted STG. 

\begin{table}[b]
\scriptsize
\centering
\caption{The Effectiveness of Variants of SCRAMBLE in case of FSM/Sequential/Scan Obfuscation.}
\label{compare}
\setlength\tabcolsep{3pt} 
\begin{tabular}{@{} l *{7}c @{}}
\toprule 
Variants  & & & \multicolumn{2}{c}{SCRAMBLE-C} & \multicolumn{2}{c}{SCRAMBLE-L} \\
\cmidrule(r){1-1}
\cmidrule(r){4-5}
\cmidrule(r){6-7}
Attacks & & & 2-stage & UB-SAT \emph{or} BMC & 2-stage & UB-SAT \emph{or} BMC\\
\cmidrule(r){1-1}
\cmidrule(lr){4-4}
\cmidrule(lr){5-5}
\cmidrule(lr){6-6}
\cmidrule(lr){7-7}

FSM & & & \cmark & \cmark & \cmark & \cmark  \\ 
Sequential Datapath & & & N/A & \cmark & N/A & \cmark$^*$  \\
Scan-chain & & & N/A & \cmark & N/A & \cmark$^*$  \\
\bottomrule
\multicolumn{7}{l}{$*$:  Requires large augmentation model incurring area overhead.} \\
\end{tabular}
\end{table}

\begin{figure}[t]
\centering
\includegraphics[width = 240pt]{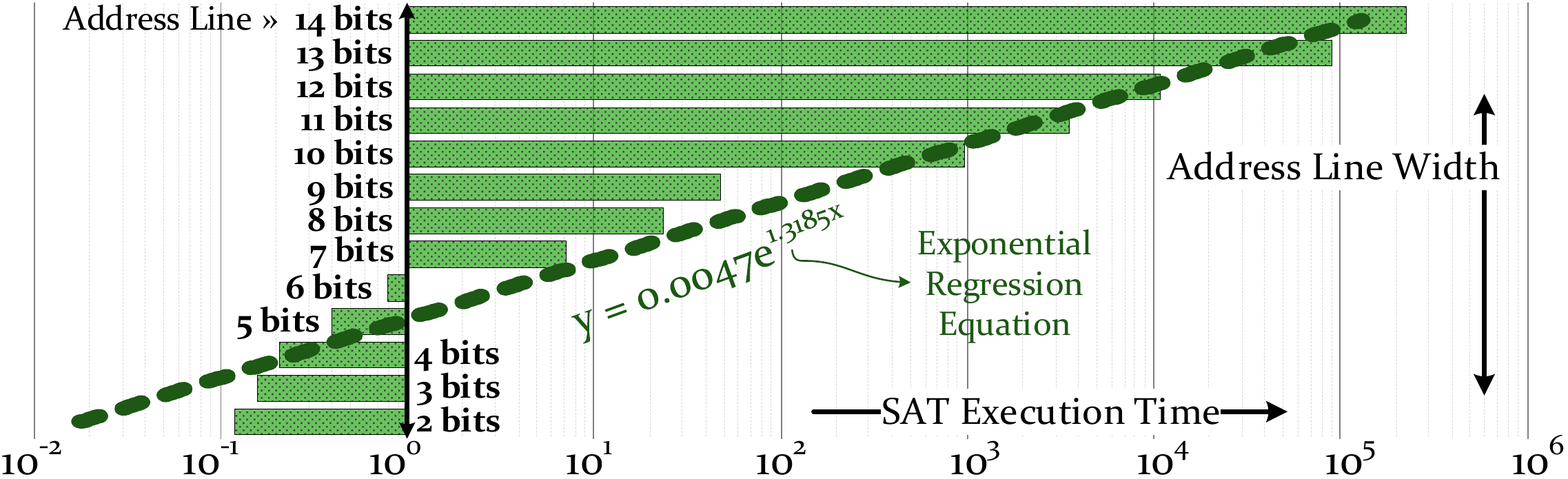}
\caption{The Impact of Address Width of the Memory on the Execution Time of its LUT-based Equivalent Model.}
\label{lutmodel}
\end{figure}

Although SCRAMBLE-L protects the design against 2-stage attacks by hiding the logic within memory, the adversary can generate the equivalent logic of the memory ($X$ input (address width) and $Y$ outputs (word size)) by replacing it with $Y$ of $X$-input LUTs, which is a fully configurable logic, and then using SAT attack. However, increasing the input size of the LUTs exponentially increases the run time of BDD-based or SAT-based attacks. Fig. \ref{lutmodel} shows that by increasing the address width (from 2 bits to 14 bits), when we replace the memory with the same size LUTs, SAT execution time increases exponentially \cite{kamali2018lut}. In addition, due to the unrolling structure of UB-SAT or BMC, these LUTs must be replicated per each iteration (per each unrolling), which makes them almost unresolvable by SAT-driven attacks. We demonstrate that UB-SAT or BMC cannot reveal the correct functionality of a design even while SCRAMBLE-L has been used with only one 256 words (address width is 8) memory.

\section{Experimental Results}

We evaluated the strength of SCRAMBLE on two sets of benchmark circuits: (1) sequential ISCAS-89 benchmark circuits and (2) few well-known small-scale ASICs to large-scale microprocessors. We have deployed a 2-stage attack according to Algorithm \ref{topofunc} to assess the strength of SCRAMBLE in FSM locking. For sequential datapath locking, we deployed an integrated BMC with SAT \cite{el2017reverse} accelerated using stages described in KC2 \cite{shamsi2019kc2}. Finally, to assess the effectiveness of scan chain locking, we implemented the ScanSAT as described in \cite{alrahis2019scansat}. All attacks are carried on a Dell PowerEdgeR620 with Intel Xeon E5-2670 2.50GHz and 64GB of RAM.

Table \ref{scrambleC} captures the execution time of scanSAT \cite{alrahis2019scansat} (for scan locking) and accelerated BMC \cite{el2017reverse, shamsi2019kc2} (for sequential datapath locking) while SCRAMBLE-C is used on ISCAS-89 benchmarks. The maximum runtime of attack is set to $10^6$ seconds, and attack will time-out (\xmark~in tables) if attack time exceeds the limit. Note that in some cases, the number of required FFs is limited. For instance, in \emph{s1196}, with 18 FFs, the maximum possible size of CRLB is only 16. As illustrated, by utilizing the CRLB with size 16, for almost all benchmark circuits, both attacks cannot retrieve the keys. Also, Table  \ref{scrambleC} reports the power, performance, and area (PPA) overhead of SCRAMBLE-C with a CRLB of size 16. While the CRLB size is fixed, the area overhead is constant and the percentage area overhead reduces when the size of the benchmark circuits increases. As shown, for even mid-size ISCAS-89 benchmark circuits, the area overhead is less than 10\%.  

\begin{table}[b]
\scriptsize 
\centering
\caption{The attack time for breaking SCRAMBLE-C used for scan chain locking and sequential datapath locking of ISCAS-89 benchmarks.}
\label{scrambleC}
\setlength\tabcolsep{1.5pt} 
\begin{tabular}{@{} l *{13}c @{}}
\toprule 
 & & & & \multicolumn{6}{c}{Attack Time (second)} & \multicolumn{3}{c}{Datapath Locking} \\ 
 \cmidrule(lr){5-10} \cmidrule(lr){11-13}
 &  &  &  & \multicolumn{3}{c}{scanSAT} & \multicolumn{3}{c}{BMC} & \multicolumn{3}{c}{PPA overhead of} \\
 \cmidrule(lr){5-7}
\cmidrule(lr){8-10}
 &  &  &  & \multicolumn{3}{c}{CRLB Size} & \multicolumn{3}{c}{CRLB Size} & \multicolumn{3}{c}{16-input CRLB } \\
\cmidrule(lr){5-7}
\cmidrule(lr){8-10}
\cmidrule(lr){11-13}
Circuit & \#FF & \#Gate & In/Out & 8 & 16 & 32 & 8 &  16 & 32 & Power & Delay & Area\\
\cmidrule(r){1-1}
\cmidrule(lr){2-4}
\cmidrule(lr){5-7}
\cmidrule(lr){8-10}
\cmidrule(lr){11-13}
s1196 & 18 & 529 & 14/14 & 2029 & \textbf{\xmark}  & \textbf{N/A}  &  1109 & \textbf{\xmark}  & \textbf{N/A} & 26.3\% & 36.5\%  & 24.1\% \\
s1423 & 74 & 657 & 17/5 & 3441 & \textbf{\xmark}  &\textbf{\xmark}  & 438.6 & 9356 & \textbf{\xmark} & 25.8\% & 28.1\%  & 23\% \\
s5378 & 179 & 2779 & 35/49 & 6406 & \textbf{\xmark}  &\textbf{\xmark}  & 6921 & \textbf{\xmark}  & \textbf{\xmark} & 8.9\% & 18.5\%  & 7.1\% \\
s9234  & 211 & 5597 & 36/39 & 1801 &\textbf{\xmark}  & \textbf{\xmark}  & 1548 & \textbf{\xmark}  & \textbf{\xmark} & 5.1\% & 14.8\%  & 3.9\% \\
s15850  & 534 & 9772 & 77/150 & 5882 &\textbf{\xmark}  & \textbf{\xmark}  & 7097 & \textbf{\xmark}  & \textbf{\xmark} & 3.1\% & 12.9\% & 2.4\% \\
s35932 & 1728 & 16065 & 35/320 & 8604 &\textbf{\xmark}  & \textbf{\xmark}  & 7110 & \textbf{\xmark}  & \textbf{\xmark} & 1.1\% & 6.5\%  & 0.9\% \\
s38584  & 1426 &  19253 & 38/304 & 4072 &\textbf{\xmark}  & \textbf{\xmark}  & 6287 & \textbf{\xmark}  & \textbf{\xmark} & 1.2\% & 5.7\% & 0.9\% \\

\bottomrule
\multicolumn{10}{l}{\textit{\textbf{\xmark} : Timeout = $10^6$ Seconds}} \\
\end{tabular}
\end{table}

To assess the robustness of SCRAMBLE for FSM locking, both SCRAMBLE-C and SCRAMBLE-L have been used on the second group of circuits. Also, the locked circuits have been evaluated using both BMC and 2-stage attacks. As illustrated in Table \ref{fsmexe}, BMC can break SCRAMBLE-C while the CRLB size is up to 16. However, for none of the circuits, BMC cannot retrieve the correct key while the CRLB size is 32. Also, in case of BMC, only utilizing a memory with 256 words (address width = 8) is enough to make the locked circuit resilient against BMC. 

Unlike BMC, which can break SCRAMBLE for small-size CRLBs and memories, 2-stage attacks are far weaker. As shown in Fig. \ref{fig-scramble-p}, since the number of false paths is extremely larger, after re-drawing the FSM using 2-stage, there is no chance for the adversary to extract the original part of the FSMs. Hence, as shown in Table \ref{fsmexe}, 2-stage attacks completely fail against SCRAMBLE.

\begin{table}[t]
\scriptsize 
\centering
\caption{The BMC and 2-stage attack time for breaking SCRAMBLE-C and SCRAMBLE-L used for FSM locking.}
\label{fsmexe}
\setlength\tabcolsep{1pt} 
\begin{tabular}{@{} l *{15}c @{}}
\toprule 
 & & & \multicolumn{12}{c}{Attack Time (second)}  \\ 
 \cmidrule(lr){4-16}
 & &  & \multicolumn{6}{c}{BMC} & \multicolumn{6}{c}{2-stage} \\
\cmidrule(lr){4-9}
\cmidrule(lr){10-15}
 & &  & \multicolumn{3}{c}{SCRAMBLE-C} & \multicolumn{3}{c}{SCRAMBLE-L} & \multicolumn{3}{c}{SCRAMBLE-C} & \multicolumn{3}{c}{SCRAMBLE-L} \\
 & &  &  \multicolumn{3}{c}{CRLB Size} & \multicolumn{3}{c}{Mem Addr} & \multicolumn{3}{c}{CRLB Size} & \multicolumn{3}{c}{Mem Addr} \\
\cmidrule(lr){4-6}
\cmidrule(lr){7-9}
\cmidrule(lr){10-12}
\cmidrule(lr){13-15}
Circuit & \#FF & \#Gate & 8 & 16 & 32 & 7 & 8 & 9 & 8 & 16 & 32 & 7 & 8 & 9 &\\
\cmidrule(r){1-1}
\cmidrule(lr){2-2}
\cmidrule(lr){3-3}
\cmidrule(lr){4-6}
\cmidrule(lr){7-9}
\cmidrule(lr){10-12}
\cmidrule(lr){13-15}
RS232 & 168 & 59 & 2.7 & 2029 & \textbf{\xmark}  &  35.7  & \textbf{\xmark}  & \textbf{\xmark} &  \textcolor{white}{...}\textbf{\xmark}\textcolor{white}{...} & \textbf{\xmark} & \textbf{\xmark} &  \textcolor{white}{...}\textbf{\xmark}\textcolor{white}{...} & \textbf{\xmark} & \textbf{\xmark} \\
32b RSA & 555 & 2139 & 1.4 & 3441 & \textbf{\xmark}  & 45.8  & \textbf{\xmark}  & \textbf{\xmark} & \textcolor{white}{..}\textbf{\xmark}\textcolor{white}{..} & \textbf{\xmark} & \textbf{\xmark} & \textbf{\xmark} & \textbf{\xmark} & \textbf{\xmark}\\
MC8051 & 578 &  6590 & 47.7 & 6406 & \textbf{\xmark}  & 50.1  & \textbf{\xmark}  & \textbf{\xmark} & \textbf{\xmark} & \textbf{\xmark} & \textbf{\xmark} & \textbf{\xmark} & \textbf{\xmark} & \textbf{\xmark}\\
SPARC  & 120K & 233K & 938 & \textbf{\xmark} & \textbf{\xmark} & 288.2  &  \textbf{\xmark}  & \textbf{\xmark} & \textbf{\xmark} & \textbf{\xmark} & \textbf{\xmark} & \textbf{\xmark} & \textbf{\xmark} & \textbf{\xmark} \\

\bottomrule
\multicolumn{10}{l}{\textit{\textbf{\xmark} : Timeout = $10^6$ Seconds}} \\
\end{tabular}
\end{table}

Since the minimum size of CRLB in SCRAMBLE-C (memory in SCRAMBLE-L), which provides a resilient FSM locking against BMC, is 32 (256 words), we reported the PPA overhead of these sizes for second groups of the circuits in Table \ref{ppafsm}. As shown, even for mid-size \emph{32b RSA} circuit, the overhead is less than 5\%. Also, the impact of increasing the size of CRLB or memory on the PPA overhead for different sizes has been illustrated in Table \ref{dcreport}. As shown, in both SCRAMBLE-C and SCRAMBLE-L, increasing either the size or address width, approximately doubles the overhead. However, compared to ISCAS-89 benchmark circuits, such as \emph{s15850} or \emph{s38584}, the incurred overhead is reasonable.

\begin{table}[t]
\scriptsize
\centering
\caption{The PPA overhead for building a locked FSM resistant to BMC attack using SCRAMBLE-C and SCRAMBLE-L}
\label{ppafsm}
\setlength\tabcolsep{1.5pt} 
\begin{tabular}{@{} l *{9}c @{}}
\toprule
& \multicolumn{4}{c}{\textbf{SCRAMBLE-C}} & \multicolumn{4}{c}{\textbf{SCRAMBLE-L}} \\
 & \multicolumn{4}{c}{(Resilient with CRLB Size = 32)} & \multicolumn{4}{c}{(Resilient with SRAM Size = $2^{8}\times$8)} \\
\cmidrule(r){2-5}
\cmidrule(r){6-9}
Circuit & RS232 & 32b RSA & MC8051 & SPARC &  RS232 & 32b RSA & MC8051 & SPARC \\
\cmidrule(r){1-1}
\cmidrule(r){2-5}
\cmidrule(r){6-9}
Area (\%) & 38.5\% & 4.5\% & 1.2\% & 0.05\% & 173\% & 17.8\% & 5.1\% & 0.1\% \\
\cmidrule(r){1-1}
\cmidrule(r){2-5}
\cmidrule(r){6-9}
Power (\%) & 44.8\%  & 5.6\%  & 1.7\%  & 0.1\%  & 224\%  & 26.8\%  & 7.2\%  & 0.3\%  \\
\cmidrule(r){1-1}
\cmidrule(r){2-5}
\cmidrule(r){6-9}
Delay (\%) & 48.4\% & 10.8\%  & 11.4\%  & 9.7\%  & 22.7\%  & 5.5\%  & 6.8\%  & 3.9\%  \\
\bottomrule
\end{tabular}
\end{table}

\begin{table}[t]
\scriptsize 
\centering
\caption{The PPA overhead of SCRAMBLE-C and SCRAMBLE-L when using CRLB/SRAM of different sizes.}
\label{dcreport}
\setlength\tabcolsep{1.5pt} 
\begin{tabular}{@{} l *{10}c @{}}
\toprule
& \multicolumn{4}{c}{\textbf{SCRAMBLE-C}} & \multicolumn{3}{c}{\textbf{SCRAMBLE-L}} & \multicolumn{2}{c}{Sample ISCAS-89} \\
 & \multicolumn{4}{c}{(CRLB Input Size)} & \multicolumn{3}{c}{(SRAM Size)} & \multicolumn{2}{c}{Benchmarks} \\
\cmidrule(r){2-5}
\cmidrule(r){6-8}
\cmidrule(r){9-10}
Overhead & 8 & 16 & 32 & 64 & ($2^{7}\times$8) & ($2^{8}\times$8) & ($2^{9}\times$8) & s15850 & s38584 \\
\cmidrule(r){1-1}
\cmidrule(r){2-5}
\cmidrule(r){6-8}
\cmidrule(r){9-10}
Area ($um^2$) &  58.1 & 136.7 & 330.8 & 620.4 & 305.8 & 612.1 & 1119 & 6262 & 21458 \\
\cmidrule(r){1-1}
\cmidrule(r){2-5}
\cmidrule(r){6-8}
\cmidrule(r){9-10}
Power ($uW$) & 4.5 & 6.9 & 14.5 & 31.9 & 31.4 & 80.6 & 118.9 & 325.7 & 1031 \\
\cmidrule(r){1-1}
\cmidrule(r){2-5}
\cmidrule(r){6-8}
\cmidrule(r){9-10}
Delay ($ns$) & 0.33 & 0.37 & 0.48 & 0.56 & 0.17 & 0.18 & 0.19 & 1.24 & 2.68 \\
\bottomrule
\end{tabular}
\end{table}

\section{Conclusion}

In this paper, we introduce SCRAMBLE, as a comprehensive obfuscation solution for protecting FSMs, sequential circuits, and scan chains against IP piracy and reverse engineering. The proposed solution, SCRAMBLE, resist against both (1) the 2-stage attacks on FSM, and (2) unrolling-based SAT attacks while sequential or scan obfuscation is targeted. We have discussed two variants of SCRAMBLE: (a) SCRAMBLE-C, and (b) SCRAMBLE-L. The SCRAMBLE-C relies on the SAT-hard and key-controlled modules that are constructed using the near non-blocking logarithmic switching network. The SCRAMBLE-L uses input multiplexing techniques to hide a part of the FSM in a memory. In our result section, we illustrated that attack time could be made unreasonably long using any of these techniques.   

\section{Acknowledgement} \label{Acknowledgement}

This research is supported by the Defense Advanced Research Projects Agency (DARPA-AFRL, \#FA8650-18-1-7819) and National Science Foundation(NSF, \#1718434).

\renewcommand{\IEEEbibitemsep}{0pt plus 0.5pt}
\makeatletter
\IEEEtriggercmd{\reset@font\normalfont\fontsize{7.0pt}{8pt}\selectfont}
\makeatother
\IEEEtriggeratref{1}

\bibliographystyle{IEEEtran}
\bibliography{IEEEabrv,refs}

\end{document}